\documentstyle[12pt]{article}

\def\bs{\begin{subequations}}
\def\es{\end{subequations}}

\catcode`\@=11

\newtoks\@stequation
\def\subequations{\refstepcounter{equation}
  \edef\@savedequation{\the\c@equation}%
  \@stequation=\expandafter{\theequation}%   %only want \theequation
  \edef\@savedtheequation{\the\@stequation}% % expanded once
  \edef\oldtheequation{\theequation}%
  \setcounter{equation}{0}%
  \def\theequation{\oldtheequation\alph{equation}}}

\def\endsubequations{\setcounter{equation}{\@savedequation}%
  \@stequation=\expandafter{\@savedtheequation}%
  \edef\theequation{\the\@stequation}\global\@ignoretrue}
\catcode`\@=12

\catcode`\@=11
\def\vereq#1#2{\lower3pt\vbox{\baselineskip1.5pt \lineskip1.5pt
\ialign{$\m@th#1\hfill##\hfil$\crcr#2\crcr\sim\crcr}}}
\catcode`\@=12

\makeatletter%  allow access to internal LaTeX commands
        \renewcommand{\theequation}{\thesection.\arabic{equation}}%
        \@addtoreset{equation}{section}%
\makeatother%  turn off access to internal LaTeX commands

\renewcommand{\thefootnote}{\fnsymbol{footnote}}

\def\dd{_{\textbf{\raisebox{.3ex}{.}}}} %dot subscript
\def\la{\langle 0 \mid}
\def\ra{\mid 0 \rangle}

\begin{document}
\begin{titlepage}
\begin{center}
March 15, 1999           \hfill UCB-PTH-99/05    \\
To appear in IJMPA              \hfill LBNL-42937    \\
                                \hfill hep-th/9903131    \\

\vskip .25in

{\large \bf Infinite Dimensional Free Algebra \\ and the 
Forms of the Master Field \\}

\vskip 0.3in

M.B. Halpern$^{a,b}$\footnote{E-mail: halpern@physics.berkeley.edu}
 and C. Schwartz$^a$

\vskip 0.15in

$^a${\em Department of Physics,
     University of California\\
     Berkeley, California 94720}\\
and\\
$^b${\em Theoretical Physics Group\\
     Ernest Orlando Lawrence Berkeley National Laboratory\\
     University of California,
     Berkeley, California 94720}
        
\end{center}

\vskip .3in

\vfill

\begin{abstract}
We find an infinite dimensional free algebra which lives at large N 
in any SU(N)-invariant action or Hamiltonian theory of bosonic matrices.  
The natural basis of this algebra is a free-algebraic generalization of 
Chebyshev polynomials and the dual basis is closely related to the planar 
connected parts.  This leads to a number of free-algebraic forms of the 
master field including an algebraic derivation of the Gopakumar-Gross form. 
For action theories,  these forms of the master field immediately give a
number of new free-algebraic packagings of the planar Schwinger-Dyson equations.
\end{abstract}

\vfill

\end{titlepage}

\renewcommand{\thefootnote}{\arabic{footnote}}
\setcounter{footnote}{0}
\renewcommand{\thepage}{\arabic{page}}
\setcounter{page}{1}

\section{Introduction}%1

Recently~\cite{us}, we have studied the algebras of phase-space master fields in
general matrix models, obtaining in particular a number of new free
algebras which generalize the Cuntz algebra. Among these generalizations,
our starting point in this paper is the set of  \emph{interacting Cuntz
algebras}~\footnote{Certain powers of $\sqrt{2}$ are scaled out here 
relative to the operators $A$ and $C$ of Ref.~\cite{us}.}
\bs \label{a}
\begin{equation}
B_{m} = {\sqrt 2} A_{m} = F_{m}(\phi) + i \pi_{m}, \;\;\;\;\; B^{\dagger}_{m} 
= F_{m}(\phi)-i\pi_{m},\;\;\;\;\; E_{mn}(\phi) = 2 C_{mn}(\phi)
\end{equation}
\begin{equation}
B_{m}B^{\dagger}_{n} = E_{mn}
\end{equation}
\begin{equation}
B^{\dagger}_{m}(E^{-1})_{mn}B_{n} 
= 1 - \ra \la 
\end{equation}
\begin{equation}
B_{m} \ra = \la B^{\dagger}_{m} = 0, \;\;\;\;\;\; m,n=1 \ldots d 
\end{equation}
\es 
which occur at large N in general bosonic matrix models, and may also occur in
matrix  models with fermions. The fields $\phi_{m}$ and $\pi_{m}$
 are the master field and the reduced momenta respectively,
and the operators $F_{m}$ and $E_{mn}$ are determined by the potential. The Cuntz
algebra is the special case of (\ref{a}) obtained in the case of  matrix
oscillators.

In the present paper, we will generalize these algebras in two
directions. First, we recall that  the  ``fifth-time'' formulation 
(see for example Ref.~\cite{Green}) maps any Euclidean action theory into a
 higher-dimensional
theory~\footnote{In an evident parallel with the AdS/CFT 
correspondence [3-5], the
fifth-time formulation also gives the  large N action theory as a classical
solution of the higher-dimensional theory (see Subsec.~2.1 and App.~A)} with 
a Hamiltonian formulation. This allows us to read Ref.~\cite{us} as 
 a unified free-algebraic treatment of action and phase-space master
fields (see Sec.~2). The unified formulation  includes and extends 
Haan's~\cite{Haan} 
early free-algebraic formulation of action master fields, and one  sees in
particular that the interacting Cuntz algebra (\ref{a}) occurs in the same way for
action and phase-space master fields. The operators  $F_{m}$ and $E_{mn}$ of the
algebra (\ref{a}) are straightforward to compute explicitly  for the
action case.

The second direction is the main subject of this paper. For action
and/or phase-space master fields, the interacting Cuntz algebra can be
extended to an infinite dimensional free algebra (see Secs.~3, 4, 5 
and 6),
whose structure, especially in the action case, controls the large N
theory. The annihilation operators of this algebra are defined as
composites of the interacting Cuntz operators
\bs \label{ez}
\begin{equation}
B_{w} = B^{w} = B_{m_{1}} \ldots B_{m_{n}}, \;\;\;\;\;\; w=m_{1} \ldots m_{n}, 
\;\;\;\;\;\; [w]=n \label{f}
\end{equation}
\begin{equation}
B_{\bar{w}} = B^{\bar{w}} = B_{m_{n}} \ldots B_{m_{1}}, \;\;\;\;\;\; 
\bar{w}=m_{n} \ldots m_{1}, \;\;\;\;\;\; [\bar{w}]=n 
\end{equation}
\es
where $w$ is any word, composed of letters $m_{i}$, and $[w]$ is 
the length of 
$w$. When a word $w$ is written as a subscript it is a label but when 
written as a superscript it is to be read as an exponent, 
producing an ordered product, as in (\ref{f}). The null word is $0$, 
with $[0]=0$ and $B^{0}=1$.  This word notation, and 
the rule
\begin{equation}
ww^{\prime} = m_{1} \ldots m_{n} m^{\prime}_{1} \ldots 
m^{\prime}_{n^{\prime}}, \;\;\;\;\;\; [ww^{\prime}] = [w]+[w^{\prime}]
\end{equation}
will be followed uniformly below.

Surprisingly, the composite annihilation operators (\ref{ez}) and the 
corresponding creation 
operators turn out to be linear in the reduced momenta $\pi_{m}$,
\begin{equation}
B_{mw} = F_{mw}(\phi) + i \pi_{m} G_{w}(\phi), \;\;\;\;\;\; 
B_{mw}^{\;\;\;\;\;\;\dagger} = F_{mw}(\phi)^{\dagger} -i 
G_{\bar{w}}(\phi) \pi_{m} \label{bq}
\end{equation}
 and this fact underlies the simple form of the infinite dimensional 
free algebra below. The operators $G_{w}$ and $F_{w}$ can be obtained in terms 
of $F_{m}$ and $E_{mn}$ (see Sec.3), and $G_{w}$, $F_{w}$ turn out to be 
free-algebraic
generalizations of Chebyshev polynomials (see Subsecs.~3.2 and 3.3).

The  infinite dimensional free algebra is then
\bs \label{c}
\begin{equation}
B_{w}B_{w^{\prime}}= B_{ww^{\prime}}, \;\;\;\;\;\; B_{w}^{\;\;\;
\dagger}B_{w^{\prime}}^{\;\;\;\dagger}
= B_{w^{\prime}w}^{\;\;\;\;\;\;\dagger}
\end{equation}
\begin{equation}
B_{mw}^{\;\;\;\;\;\;\dagger}B_{nw^{\prime}} = G_{\bar{w}m}B_{nw^{\prime}}-
G_{\bar{w}}B_{mnw^{\prime}} \label{bf}
\end{equation}
\begin{equation}
B_{mw}^{\;\;\;\;\;\dagger}(E^{-1})_{mn}B_{nw^{\prime}} =  B_{w}^{\;\;
\dagger}G_{w^{\prime}} + 
G_{\bar{w}}B_{w^{\prime}} - G_{\bar{w}w^{\prime}}
 - G_{\bar{w}} \ra \la G_{w^{\prime}} \label{b}
\end{equation}
\begin{equation}
B_{mw}B_{nw^{\prime}}^{\;\;\;\;\;\;\dagger} = \sum_{w^{\prime\prime}} 
(B_{mw^{\prime\prime}}f_{w^{\prime\prime},w,n,w^{\prime}} + 
f^{*}_{w^{\prime\prime},w^{\prime},m,w}B_{nw^{\prime\prime}}^{\;\;\;\;\;\;
\dagger}) + E_{mw;nw^{\prime}}(\phi) \label{bc}
\end{equation} 
\begin{equation}
B_{w} \ra = \delta_{w,0}\ra, \;\;\;\;\;\; \la B_{w}^{\;\;\;\dagger} = 
\la \delta_{w,0}
\end{equation}
\es
where $E_{mw,nw^{\prime}}(\phi)$  and the structure constants $f$ will be given 
in Sec.~6. The interacting
Cuntz algebra (\ref{a}) is  a subalgebra of (\ref{c}),  and
(\ref{bf}) includes  a new relation for $B^{\dagger}_{m}B_{n}$. In the case of
 oscillators
and/or free action theories, the Cuntz algebra itself is a subalgebra  of
the infinite dimensional algebra (see App.~B).

The creation operators of this algebra provide us with a natural basis
\begin{equation}
B_{w} ^{\;\;\;\dagger} \ra = G_{\bar{w}}(\phi) \ra \label{d}
\end{equation}
comprised of the $G_{w}$'s themselves, and the dual basis, orthonormal to 
(\ref{d}), turns out to involve the
planar connected parts $X_{w}$ in a very simple way. 

This leads us to a number of forms of
the master field (see Sec.~7), including the basic form
\begin{equation}
\phi_{m} = \sum_{w} X_{mw} G_{\bar{w}}(\phi) \label{ef}
\end{equation}
and the dual basis form
\begin{equation}
\phi_{m} = \phi_{m}^{\dagger} = \bar{B}_{m}(1+\bar{X}(B^{\dagger})), 
\;\;\;\;\;\; \bar{B}_{m} = (E^{-1})_{mn}B_{n} \label{ed}
\end{equation}
where $\bar{B}$ and $B^{\dagger}$ satisfy a Cuntz algebra and 
$\bar{X}(B^{\dagger})$ is a 
 generating function of planar connected parts.  
The dual basis form (\ref{ed}) is the hermitian counterpart of the 
non-hermitian form obtained diagrammatically by Gopakumar and 
Gross~\cite{Gopa}.  We also give the forms of the master field in 
terms of the planar correlators and the planar 1PI parts.

For action theories, these forms of the master field immediately give 
a number of new free-algebraic forms (see Sec.~8) of the planar Schwinger-Dyson 
equations, including, surprisingly, the basic form (\ref{ef}) itself and the 
dual basis system
\begin{equation}
B^{\dagger}_{m} + E_{mn}(\bar{B}(1+\bar{X}))\bar{B}_{n} = 
G_{m}(\bar{B}(1+\bar{X})) \label{es}
\end{equation}
both of which can be used for computation of
 the planar connected parts. Systems similar to (\ref{es}) follow for the planar 
correlators and the planar effective action, and, 
although they are packaged differently, these systems (including 
(\ref{es})) are closely related 
to the free-algebraic equations derived diagrammatically in 
Ref.~\cite{Cvit}. 

We conclude that the interacting Cuntz algebra (\ref{a}) and the 
infinite dimensional free algebra (\ref{c}) provide an algebraic 
framework which underlies and extends much of what is known about 
large N, and we are optimistic that these algebras will provide a 
foundation for the future study of the master field.

\section{Unification of Action and Phase-Space \\ Master Fields}%2

\subsection{Fifth-time formulation and Euclidean quantum \\ field 
theory}%2.1

We consider a general SU(N)-invariant matrix model with  Euclidean action $S$
\bs
\begin{eqnarray}
<Tr\phi^{w} > = \eta^{-1} \int (d\phi)\; e^{-S}\; Tr[\phi^{w}]
, \;\;\;\;\;\; \eta=\int (d\phi) e^{-S} \label{ao} \\
S = N Tr[{\cal S}(\frac{\phi}{\sqrt N})], \;\;\;\;\;\; \phi^{w} = 
\phi^{m_{1}} \ldots \phi^{m_{n}}, \;\;\;\;\;\; m=1 \ldots d
\end{eqnarray}
\es
and follow the fifth-time formulation~\cite{Green} to interpret the model as a
quantum system, with a (fifth time) Hamiltonian formulation,  in one higher 
dimension.
The resulting picture is  a pedestrian version of operator
Euclidean quantum field theory.

 In the Hamiltonian formulation, the matrix fields $\phi^{m}$ are 
 operators and the action averages are reinterpreted as ground state averages:
\begin{equation}
<Tr \phi^{w}> = \langle \dd 0 \mid Tr \phi^{w} \mid 0 \dd \rangle , 
\;\;\;\;\;\; \mid 0 \dd \rangle = \psi_{0}(\phi) = \eta^{-\frac{1}{2}} 
e^{-\frac{S}{2}} \label{i}
\end{equation}
where the dot in the (unreduced) ground state follows the notation of
Ref.~\cite{us}. We may also introduce momentum operators and equal fifth-time
commutators as
\bs
\begin{equation}
\pi^{m}_{rs} = \frac{1}{i} \frac{\partial}{\partial \phi^{m}_{sr}}, 
\;\;\;\;\;\; [\phi^{m}_{rs},\pi^{n}_{tu}] = i\; \delta^{mn}\; 
\delta_{st}\; \delta_{ru}
\end{equation}
\begin{equation}
(\phi^{m}_{rs})^{\dagger}=\phi^{m}_{sr},  \;\;\;\;\;\; 
(\pi^{m}_{rs})^{\dagger}=\pi^{m}_{sr},\;\;\;\;\;\; r,s = 1 \ldots N
\end{equation}
\es
and, following Ref.~\cite{us}, we  use the momenta to construct 
matrix creation and annihilation operators
\bs \label{g}
\begin{eqnarray}
B^{m}_{rs} = {\sqrt 2} A^{m}_{rs} &=& F^{m}_{rs} + i 
\pi^{m}_{rs},\;\;\;\;\;\; B^{m}_{rs} \mid 0 \dd \rangle = 0 \\
(B^{\dagger m})_{rs} = {\sqrt 2} (A^{\dagger m})_{rs} &=& F^{m}_{rs} - i 
\pi^{m}_{rs},\;\;\;\;\;\;\langle \dd 0 \mid (B^{\dagger m})_{rs} = 0 \\
F^{m}_{rs} &=& \frac{1}{2} \frac{\partial S}{\partial \phi^{m}_{sr}}.
\end{eqnarray}
\es
Ref.~\cite{us} also tells us that  the quantities
\begin{equation}
E^{mn}_{rs} = 2 C^{mn}_{rs} = [B^{m}_{rt},(B^{\dagger n})_{ts}] = 
\frac{\partial^{2} S}{\partial \phi^{m}_{tr} \; \partial 
\phi^{n}_{st}} \label{dp}
\end{equation}
will be useful at large N.

As for the fifth-time Hamiltonian itself, we may choose any of a very large
number of operators, for example
\begin{equation}
H_{5} = \frac{1}{2} Tr( B^{\dagger m}\; B^{m}), \;\;\;\;\;\; H_{5} 
\mid 0 \dd \rangle = 0 \label{e}
\end{equation}
so long as  the choice provides us with a healthy Hilbert space and 
its ground state is $\mid 0 \dd \rangle$ in (\ref{i}).
 The equal fifth-time averages of any such
higher-dimensional  system will be the original Euclidean action averages,
and, moreover, the large N action averages are controlled by the 
phase-space master fields~\cite{us}, which are classical solutions of the
higher-dimensional theory.
The parallel with the AdS/CFT correspondence [3-5] is clear,
 if only we are clever 
enough to choose both an interesting action theory and an interesting
higher-dimensional extension. Except for a simple
example based on (\ref{e}) in App.~A, further consideration of this issue is
beyond the scope of  the present paper, and we will not choose any specific
form for $H_{5}$ here. 

\subsection{Reduced formulation}%2.2

We may now go over to reduced \footnote{The reduced formalism 
(Wigner-Eckhart for SU(N)) was pioneered by Bardakci \cite{Bard} 
and applied to large N matrix mechanics in \cite{Halp,usold}.}
states and operators for the large N action
theory, drawing heavily on the results of Ref.~\cite{us}. Important relations
 given there include
\begin{equation}
\langle \dd 0 \mid Tr[(\frac{\phi}{\sqrt N})^{w}] \mid 0 \dd \rangle = 
N \la \phi^{w} \ra \equiv N <\phi^{w}>
\end{equation}
where $\phi_{m}$ is the master field, $\phi^{w}$ are products of the master field
in the word notation (\ref{f}), and the undotted vacuum is the reduced ground 
state.  The reduced equal (fifth) time algebra involves the 
tilde operators introduced in Ref.~\cite{us}
\bs \label{k}
\begin{eqnarray}
[\phi_{m},\tilde{\pi}_{n}] = [ \tilde{\phi}_{m},\pi_{n}] = i\; 
\delta_{m,n} \; \ra \la \\ 
{[}\phi_{m},\tilde{\phi}_{n}] = [ \pi_{m},\tilde{\pi}_{n}] =0 \\  
{[}\phi_{m},\pi_{m}] = i (d-1+\ra \la ) \\ 
\phi^{\dagger}_{m} = \phi_{m}, \;\;\;\;\;\;\pi^{\dagger}_{m} = \pi_{m} \\ 
\tilde{\phi}_{m} \ra = \phi_{m} \ra , \;\;\;\;\;\;\tilde{\pi}_{m} \ra = 
\pi_{m} \ra \label{ap}
\end{eqnarray}
\es
where the operators $\pi_{m}$ are the reduced momenta.

The reduced creation and annihilation operators corresponding to 
(\ref{g}) are
\begin{equation}
B_{m} = {\sqrt 2} A_{m} = F_{m}(\phi)+i\pi_{m}, \;\;\;\;\;\; 
B^{\dagger}_{m} = {\sqrt 2} A^{\dagger}_{m} = F_{m}(\phi)-i\pi_{m}. \label{o}
\end{equation} 
These operators satisfy the interacting Cuntz algebra~\cite{us}
\bs \label{ba}
\begin{equation}
B_{m}B^{\dagger}_{n} = E_{mn} (\phi)= 2C_{mn}(\phi) \label{h}
\end{equation}
\begin{equation}
B^{\dagger}_{m} (E^{-1})_{mn} B_{n} = 1 - \ra \la 
\end{equation}
\begin{equation}
B_{m} \ra = \la B^{\dagger}_{m} = 0 \label{l}
\end{equation}
\es
at equal (fifth) time, as well as the relations
\bs \label{eg}
\begin{equation}
[\tilde{B}_{m},B_{n}] = [\tilde{B}^{\dagger}_{m},B^{\dagger}_{n}] = 0
\end{equation}
\begin{equation}
[\tilde{\phi}_{p},B_{m} B^{\dagger}_{n}] = 0
\end{equation}
\begin{equation}
B_{m}\;B^{\dagger}_{n} \ra = 2i [\tilde{\pi}_{n},F_{m}] \ra = 2 
C_{mn}(\phi) \ra = E_{mn}(\phi) \ra \label{j}
\end{equation} 
\es
which will be useful below.

It should be noted that  Haan's~\cite{Haan} Euclidean master field relation
appears in our notation as
\begin{equation}
(F_{m} + i\tilde{\pi}_{m}) \ra = 0.
\end{equation}
Although this relation follows from (\ref{ap}), (\ref{o}) and (\ref{l}), 
the operators
 $F_{m}+i\tilde{\pi}_{m}$ do not satisfy any simple algebra.

\subsection{Sharpening a tool}%2.3

In Ref.~\cite{us}, the $BB^{\dagger}$ relation  (\ref{h}) was proven by  analysis
 of the ground state wave
function (and follows from (\ref{i}) in the action case), but a 
conjecture was offered which would give this result directly in the reduced 
operator formulation.  Here we prove this conjecture, assuming only the
completeness of the basis $\phi^{w} \ra$.

\vspace{6pt}
\noindent Theorem:
\begin{equation}
If \;\; [X,\tilde{\phi}_{m}] =  [Y,\tilde{\phi}_{m}] = 0, \;\;\; \forall 
m \;\;\; and \;\;\; X \ra = Y \ra, \;\;then \;\; X=Y. \label{m}
\end{equation}
 Proof:  Introduce the complete set of states
\begin{equation}
\mid w \rangle \equiv \phi^{w} \ra = \tilde{\phi}^{\bar{w}}\ra 
\end{equation}
and follow the steps
\begin{equation} 
X \mid w \rangle = X \tilde{\phi}^{\bar{w}}\ra = 
\tilde{\phi}^{\bar{w}}X \ra  
= \tilde{\phi}^{\bar{w}}Y \ra = Y \tilde{\phi}^{\bar{w}}\ra = Y \mid 
w \rangle .
\end{equation}
In practice, this theorem can be read as:
\bs
\begin{eqnarray}
&&[\tilde{\phi}_{m},O_{1}(\phi,\pi)] = 0, \;\; \forall m \;\;\;  
   \longrightarrow \;\;  O_{1}(\phi,\pi) = O_{2}(\phi) \\ 
&& \hspace{2.2cm} O_{1}(\phi,\pi)\ra = O_{2}(\phi) \ra \label{fa}
\end{eqnarray}
\es
where $O_{2}(\phi)$ is determined by the ground state condition (\ref{fa}). 
This is the form  conjectured in Ref.~\cite{us}.  
As a first application of this theorem, the relation (\ref{h}) of the 
interacting Cuntz algebra follows immediately from (\ref{eg}).

\subsection{Action examples}%2.4

Using Appendices C and E of Ref.~\cite{us}, and in particular the results,
\bs
\begin{equation}
(E^{mn})_{rs} = \frac{\partial^{2} S}{\partial \phi^{m}_{tr} \partial 
\phi^{n}_{st}}  \;\stackrel{_{\textstyle =}}{_{_{N}}} \;\;
 B^{m}_{rt}(B^{\dagger n})_{ts}
\end{equation}
\begin{equation}
\frac{1}{N} Tr[h(\frac{\phi}{\sqrt N})]  \;\stackrel{_{\textstyle =}}{_{_{N}}} 
\;\; <h(\phi)> \label{ce}
\end{equation}
\es
we may compute the operators $F_{m}$ and $E_{mn}$ of the interacting Cuntz 
algebra (\ref{o}),(\ref{ba}) for any \vspace{6pt} action:

\noindent 1. Standard one-matrix model.
\bs \label{et}
\begin{equation}
S = Tr (\frac{m^{2}}{2} \phi^{2} + \frac{\lambda}{4N} \phi^{4})
\end{equation}
\begin{equation}
F = \frac{1}{2} (m^{2} \phi + \lambda \phi^{3}), \;\;\;\;\;\; E = 
m^{2} + \lambda (<\phi^{2}> + <\phi>\phi + \phi^{2}). \label{by}
\end{equation}
\es
\noindent 2. General one-matrix model.
\bs
\begin{equation}
S = N \sum_{n=1}^{\infty} \frac{S_{n}}{n} Tr[(\frac{\phi}{\sqrt N})^{n}]
\end{equation}
\begin{equation}
F= \frac{1}{2} \sum_{n=1}^{\infty} S_{n} \phi^{n-1}, \;\;\;\;\;\; 
E = \sum_{n=2}^{\infty} S_{n} \; \sum_{m=0}^{n-2}\; <\phi^{m}>\; 
\phi^{n-m-2}.
\end{equation}
\es
\noindent 3. Two-matrix model.
\bs
\begin{equation}
S = Tr[\frac{m_{1}^{2}}{2} (\phi^{1})^{2} + \frac{m_{2}^{2}}{2} (\phi^{2})^{2} + 
\frac{\lambda_{1}}{4N} (\phi^{1})^{4}+\frac{\lambda_{2}}{4N} (\phi^{2})^{4} + g
 \phi^{1}\phi^{2} ]
\end{equation}
\begin{equation}
F_{1} =\frac{1}{2} (m_{1}^{2} \phi_{1} + \lambda_{1} \phi_{1}^{3} + g \phi_{2})
, \;\;\;\;\;\;
F_{2} =\frac{1}{2} (m_{2}^{2} \phi_{2} + \lambda_{2} \phi_{2}^{3} + g \phi_{1})
\end{equation}
\begin{equation}
E_{11} = m_{1}^{2} + \lambda_{1}(\phi_{1}^{2} + <\phi_{1}> 
\phi_{1} + <\phi_{1}^{2}> ), \;\;\;\;\;\; E_{12}=g
\end{equation}
\begin{equation}
E_{22} = m_{2}^{2} + \lambda_{2}(\phi_{2}^{2} + <\phi_{2}> 
\phi_{2} + <\phi_{2}^{2}> ), \;\;\;\;\;\; E_{21} =g .
\end{equation}
\es
\noindent 4. General action.
\bs
\begin{equation}
S = N \sum_{w} S_{w} Tr[(\frac{\phi}{\sqrt N})^{w}]
\end{equation}
\begin{equation}
F_{m} = \frac{1}{2} \sum_{w} S_{w} \sum_{w=umv} \phi^{vu}
, \;\;\;\;\;\;
E_{mn} = \sum_{w} S_{w} \sum_{w \sim numv} <\phi^{u}> \phi^{v}
\end{equation}
\es
where the notation $w \sim w^{\prime}$ means that the two words are equivalent 
under a cyclic permutation of their letters.

For actions with even powers of $\phi$ only, we may set the odd vev's to zero.
We also find that the simple forms
\begin{equation}
F_{m} = F_{m}(\phi_{m}), \;\;\;\;\;\; E_{mn} = E_{m}(\phi_{m})\; \delta_{m,n}
 \label{fc}
\end{equation}
follow for matrix models of independent matrices (free random 
variables~\cite{Voic}). 
The special case of free actions and/or oscillators (which give the 
Cuntz algebra) is discussed in App.~B.

\section{Annihilation Operators}%3

In Secs.~3-7 below, action and phase-space master fields are discussed 
on an equal footing.

\subsection{Linear in $\pi$}%3.1

We turn now to the construction of the infinite dimensional free
algebra, beginning with
the composite annihilation operators $B_{w}$:
\begin{equation}
B_{w} \equiv B^{w} = B_{m_{1}} B_{m_{2}} \ldots B_{m_{n}}, 
\;\;\;\;\;\; B_{w} \ra = \delta_{w,0} \ra.
\end{equation}
These operators automatically satisfy the product rule
\begin{equation}
B_{w} B_{w^{\prime}} = B_{ww^{\prime}} \label{q}
\end{equation}
and moreover we find with (\ref{k}) and (\ref{l}) that
\bs
\begin{equation}
[\tilde{\phi}_{p},B_{m}] = - \delta_{p,m} \ra \la \label{ew}
\end{equation}
\begin{equation}
[\tilde{\phi}_{p},B_{mn}] = - \delta_{p,m} \ra \la B_{n} = - 
\delta_{p,m} \ra \la 2 F_{n}(\phi)
\end{equation}
\begin{equation}
\la \xi (\phi) \pi_{m} = \la \{ [\tilde{\xi}(\phi), \pi_{m}] -i 
F_{m}(\phi) \tilde{\xi}(\phi) \} = \la \xi_{m}(\phi) 
\end{equation}
\es
where the operators $\xi_{m}(\phi)$ are determined in principle as in 
Ref.~\cite{us}. It
follows  that
\begin{equation}
[\tilde{\phi}_{p},B_{mw}] = - \delta_{p,m} \ra \la B_{w} 
= - \delta_{p,m} \ra \la G_{w}(\phi)
\end{equation}
where the operators $G_{w}$ are to be determined. The theorem in 
(\ref{m}) then
tells us that the annihilation operators are linear in the reduced
momenta $\pi_{m}$
\bs \label{cu}
\begin{equation}
B_{mw} = F_{mw}(\phi) + i \pi_{m} G_{w}(\phi) \label{n}
\end{equation}
\begin{equation}
 G_{0} = 1, \;\;\;\;\;\; G_{m} = 2F_{m}
\end{equation}
\es
where the operators $F_{w}$ are also to be determined. In what follows, we will
discuss this surprising result from a number of viewpoints.

\subsection{Determination of $F_{w}$ and $G_{w}$}%3.2

In this subsection, we give an independent inductive proof of the 
formula (\ref{n}) which also
determines the coefficients $F_{w}$ and $G_{w}$ recursively in terms of the known
operators $F_{m}$ and $E_{mn}$.

To begin, we rewrite the interacting Cuntz relation (\ref{h}) in terms of 
reduced momenta, using (\ref{o}):
\begin{equation}
B_{m} B^{\dagger}_{n} = E_{mn}(\phi) \longleftrightarrow \pi_{m}\pi_{n} + 
i\pi_{m}F_{n} -i F_{m}\pi_{n} +F_{m}F_{n} - E_{mn} = 0. \label{r}
\end{equation}
The $\pi$ form of this relation will be called the \emph{first master
constraint} below. It allows us to eliminate $\pi_{m} \pi_{n}$ in
favor of terms linear in $\pi$, and hence to verify for example that
$B_{mn}=B_{m}B_{n}$ is indeed linear in $\pi$. A proof by induction 
then starts with 
\begin{equation}
B_{m} B_{nw} = B_{mnw} \longleftrightarrow (F_{m} + i\pi_{m})(F_{nw} + i
 \pi_{n}G_{w}) = F_{mnw} + i \pi_{m} G_{nw} \label{p}
\end{equation}
where we have assumed the form (\ref{n}) and the left side of (\ref{p})
 is a special case of (\ref{q}).

Using (\ref{r}) in (\ref{p}), one then obtains the recursion relations
\bs
\begin{equation}
G_{mw} = F_{mw} + F_{m}G_{w}
\end{equation}
\begin{equation}
F_{mnw} = F_{m}(F_{nw} + F_{n}G_{w}) - E_{mn}G_{w}
\end{equation}
\es
which can be rearranged into the more useful forms
\bs \label{cv}
\begin{equation}
G_{mnw} = G_{m}G_{nw} -E_{mn}G_{w} \label{bx}
\end{equation}
\begin{equation}
F_{mw} = G_{mw} - F_{m}G_{w}. \label{ad}
\end{equation}
\es
These relations are easily iterated to any desired order, and we list here
the results
\bs
\begin{equation}
G_{0} = 1, \;\;\;\;\;\;G_{m} = 2F_{m}, \;\;\;\;\;\; 
G_{mn}=G_{m}G_{n}-E_{mn}
\end{equation}
\begin{equation}
G_{mnp} = G_{m}G_{n}G_{p}-G_{m}E_{np}-E_{mn}G_{p}
\end{equation}
\begin{equation}
G_{mnpq} = G_{m}G_{n}G_{p}G_{q} - G_{m}G_{n}E_{pq}-G_{m}E_{np}G_{q} - 
E_{mn}G_{p}G_{q} + E_{mn}E_{pq}
\end{equation}
\begin{equation}
F_{mn} = 2F_{m}F_{n}-E_{mn}, \;\;\;\;\;\; F_{mnp} = 4F_{m}F_{n}F_{p} - 
F_{m}E_{np} - 2E_{mn}F_{p}
\end{equation}
\es
for the first few words of $F$ and $G$.

More generally,  the recursion relations can be used to prove the
following properties
\begin{equation}
G_{w}^{\;\;\;\dagger} = G_{\bar{w}}, \;\;\;\;\;\; F_{mw}^{\;\;\;\;\;\dagger} = 
G_{\bar{w}m} - G_{\bar{w}}F_{m}
\end{equation}
\bs
\begin{equation}
{\cal G} = \frac{1}{1-\alpha_{m}G_{m}(\phi) + \alpha_{m} \alpha_{n} 
E_{mn} (\phi)} = \sum_{w} \alpha^{w} G_{w}(\phi) \label{s}
\end{equation}
\begin{equation}
 (1-\alpha_{m}F_{m}(\phi)) \; {\cal G} = \sum_{w} \alpha^{w} 
F_{w} (\phi), \;\;\;\;\;\; F_{0}=1 \label{t}
\end{equation}
\es
\begin{equation}
G_{wm} G_{nw^{\prime}} = G_{wmnw^{\prime}} +G_{w}E_{mn}G_{w^{\prime}}.\label{cl}
\end{equation}
Here $\alpha_{m}$ (with products $\alpha^{w}$) is a free-algebraic 
source or ``place marker''
whose only property is that it commutes with $\phi_{m}$ and 
$\pi_{m}$.

The generating functions (\ref{s}) and 
(\ref{t}) show that
$G_{w}$ and $F_{w}$ are free-algebraic generalizations of Chebyshev
polynomials (see also Subsec.~3.3 and App.~B).

We also mention the relations
\begin{equation}
B_{mw} = -B^{\dagger}_{m} G_{w} + G_{mw}, \;\;\;\;\;\; 
B_{mw}^{\;\;\;\;\;\dagger} = -G_{\bar{w}} B_{m} + G_{\bar{w}m} \label{bm}
\end{equation}
which are a useful alternative to the basic equation (\ref{n}), and
 the relations 
\bs
\begin{equation}
G_{wmn} = G_{wm}G_{n} - G_{w}E_{mn}, \;\;\;\;\;\; G_{0}=1, \;\;\;\;\;\; 
G_{m} = 2F_{m}
\end{equation}
\begin{equation}
F_{wmn} = F_{wm}G_{n} - F_{w}E_{mn}, \;\;\;\;\;\; F_{0}=1, \;\;\;\;\;\; 
F_{m} = 1F_{m}
\end{equation}
\es
which show a complete symmetry of the recursion relations for $G_{w}$ 
and $F_{w}$, except for their initial conditions.  The relations
\bs
\begin{equation}
\pi_{m}G_{w} \ra = i F_{mw} \ra, \;\;\;\;\;\; B_{m}^{\dagger} G_{w} 
\ra = G_{mw} \ra \label{ae}
\end{equation}
\begin{equation}
[i\pi_{m},\tilde{G}_{nw}]\ra = E_{mn}G_{w} \ra \label{db}
\end{equation}
\begin{equation}
[i\pi_{m},\tilde{F}_{n}]\ra = C_{mn}(\phi) \ra \label{da}
\end{equation}
\es
also follow from the discussion above. The relation (\ref{da}), which is a 
special case of (\ref{db}), was given in Ref.~\cite{us}.

\subsection{One-Matrix models}%3.3

In the case of general one-matrix (action or Hamiltonian) models the
operators $F$ and $E$ commute, and $w \rightarrow [w]$, giving the simpler forms
\bs
\begin{equation}
G_{n+2} = G_{1} G_{n+1}-EG_{n},\;\;\;\;\;\; G_{0}=1,\;\;\;\;\;\;G_{1}=2F
\end{equation}
\begin{equation}
F_{n+1}=G_{n+1}-FG_{n}, \;\;\;\;\;\; F_{0}=1, \;\;\;\;\;\; F_{1}=F
\end{equation}
\begin{equation}
G_{n} = E^{\frac{n}{2}} \frac{sin ((n+1)\theta)}{sin \theta}, 
\;\;\;\;\;\; F_{n} = E^{\frac{n}{2}} cos (n\theta) \label{u}
\end{equation}
\begin{equation}
\rho = \frac{\sqrt E}{\pi} sin \theta, \;\;\;\;\;\; cos \theta = 
\frac{F}{\sqrt E}, \;\;\;\;\;\; E = 2C = F^{2} + \pi^{2} \rho^{2} 
\label{v}
\end{equation}
\begin{equation}
G_{m}G_{n} = \sum_{k=0}^{\min_{m,n}}E^{k} G_{m+n-2k} \label{cp}
\end{equation}
\es
which include the Chebyshev polynomials themselves in (\ref{u}). 
The finite operator product expansion in (\ref{cp}) follows immediately from 
this form.  According to Ref.~\cite{us}, 
 the quantity $\rho$ in (\ref{v}) is the ground state density of the action
or Hamiltonian system.

Another special case with simplifications is that of  many 
oscillators and/ \newline or free actions (see App.~B). 

\subsection{Master constraints}%3.4

Using (\ref{n}), the composition law
\begin{equation}
B_{mw} B_{n} = B_{mwn} \label{ai}
\end{equation}
can be written out in two equivalent forms, called the \emph{master
constraints},
\bs
\begin{equation}
\pi_{m}G_{w}\pi_{n} + i \pi_{m} F_{n\bar{w}}^{\;\;\;\;\dagger} + 
F_{mw}(-i\pi_{n}) + F_{mwn} -F_{mw}F_{n} = 0 \label{aa}
\end{equation}
\begin{equation}
B_{m}^{\dagger} G_{w} B_{n} = B^{\dagger}_{m} G_{wn} + G_{mw} B_{n} - 
G_{mwn} \label{ac}
\end{equation}
\es
and (\ref{aa}) contains the first master constraint (\ref{r}) as the special case
 when $w=0$.

More generally, the form (\ref{aa}) of the master constraints allow us to
eliminate  quadratic forms $\pi_{m}G_{w}\pi_{n}$ in favor of forms linear in the
reduced momenta, and similarly for $B^{\dagger}_{m}G_{w}B_{n}$ in 
(\ref{ac}).

In Hamiltonian theories, constraints are constants of the motion and
the first master constraint, which is equivalent to 
$B_{m}B^{\dagger}_{n}-E_{mn}=0$,
was noted as a set of $d^{2}$ constants of the motion in Ref.~\cite{us}. It is 
shown in App.~C that all the higher master constraints
are in fact composites of the first master constraint, so there are no new
independent constants of the motion in this list.

\section{Creation Operators}%4

\subsection{Creation operators and the natural basis}%4.1

The creation operators of the infinite dimensional free algebra  are
defined as the hermitian
conjugates of the annihilation operators
\bs
\begin{equation}
B_{w}^{\;\;\;\dagger}  = B^{\dagger}_{m_{n}} \ldots B^{\dagger
}_{m_{1}} = B^{\dagger}_{\;\;\bar{w}}
\end{equation}
\begin{equation}
B_{mw}^{\;\;\;\;\;\;\dagger} = F_{mw}(\phi)^{\dagger} -i 
G_{\bar{w}}(\phi) \pi_{m} \label{af}
\end{equation}
\begin{equation}
\la B_{w}^{\;\;\;\dagger} = \la \delta_{w,0}
\end{equation}
\es
and therefore satisfy the product rule
\begin{equation}
B_{w}^{\;\;\;\dagger} B_{w^{\prime}}^{\;\;\;\dagger} = B_{w'w}^{\;\;\;\;\;\;\;
\dagger}.
\label{aj}
\end{equation}
The set of all these creation operators on the ground state is a natural
complete~\cite{us} basis, and we see
from  (\ref{ad}), (\ref{ae}) and (\ref{af}) that this basis can be expressed 
in terms of the polynomial $G_{w}$'s as
\bs
\begin{equation}
(B_{\bar{w}})^{\dagger} \ra = B^{\dagger\; w} \ra =
\tilde{B}^{\dagger \;\; \bar{w}} \ra =  G_{w}(\phi) 
\ra  \label{ag} 
\end{equation}
\begin{equation}
<G_{w}(\phi)> = \delta_{w,0}. \label{ag1}
\end{equation}
\es
In what follows,  the states on the right and left  of  (\ref{ag}) will be called
the \emph{natural basis} and its operator
form respectively.  Further discussion of completeness is given in Subsec.~5.4.

\subsection{$B^{\dagger}B$ relations}%4.2

Using (\ref{n}), (\ref{af}) and the first master constraint (\ref{r}), we find the
$B^{\dagger}B$ algebra
\begin{equation}
B_{mw}^{\;\;\;\;\;\; \dagger} \; B_{nw^{\prime}} = G_{\bar{w}m}\; B_{nw^{\prime}}
 - G_{\bar{w}} \; B_{mnw^{\prime}} \label{di}
\end{equation}
and the relations
\bs \label{dj}
\begin{equation}
B^{\dagger}_{m}(E^{-1})_{mn}B_{n} = 1 - \ra \la 
\end{equation}
\begin{equation}
B_{mpw}^{\;\;\;\;\;\;\; \dagger}(E^{-1})_{mn}B_{nqw^{\prime}} = G_{\bar{w}p}
B_{qw^{\prime}} - G_{\bar{w}} B_{pqw^{\prime}} -G_{\bar{w}p} \ra \la G_{qw^
{\prime}}
\end{equation}
\es
also follow immediately from the interacting Cuntz algebra and the
composition laws (\ref{ai}) and (\ref{aj}).

A more symmetric version of (\ref{di}) and (\ref{dj}) is
\bs
\begin{equation}
B_{w}^{\;\;\dagger}B_{w^{\prime}} = B_{w}^{\;\;\dagger}G_{w^{\prime}} + 
G_{\bar{w}}B_{w^{\prime}} - G_{\bar{w}w^{\prime}} \label{dk}
\end{equation}
\begin{equation}
B_{mw}^{\;\;\;\;\;\;\dagger}(E^{-1})_{mn}B_{nw^{\prime}} = B_{w}^{\;\;\;
\dagger}B_{w^{\prime}}
 - G_{\bar{w}} \ra \la G_{w^{\prime}} \label{dl}
\end{equation}
\es
where (\ref{dk}) can be used to ``linearize'' (\ref{dl}).  These forms follow 
directly from (\ref{bm}) and the interacting Cuntz algebra.

\subsection{Local and non-local}%4.3

In Ref.~\cite{us}, many reduced operators were called non-local because they 
involved arbitrarily-high powers
of the reduced momenta $\pi_{m}$, and others were called local  because they
involved no more than two powers of the reduced momenta. The results above
blur this distinction.

As an example~\cite{us}, consider the (hermitian) isotropic oscillator
Hamiltonian $H$, which may now be reexpressed in terms of the 
generators of the infinite dimensional free algebra:
\bs
\begin{eqnarray}
&&H \equiv \sum_{w \neq 0} A_{w}^{\;\;\; \dagger} A_{w} = \sum_{w \neq 0} 
\frac{1}{2^{[w]}} B_{w}^{\;\;\;\dagger}B_{w}  \\ 
&&= \sum_{m,w}\frac{1}{2^{[w]+1}}(G_{\bar{w}m} B_{mw} - 
G_{\bar{w}}B_{mmw}) \label{ak} \\
&&=  \sum_{m,w}\frac{1}{2^{[w]+1}}( B_{mw}^{\;\;\;\;\;\;\dagger}\;G_{mw} - 
B_{mmw}^{\;\;\;\;\;\;\;\;\;\dagger}\;G_{w}) .\label{al}
\end{eqnarray}
\es
The starting point is ``non-local'' because each of the Cuntz operators in
the products $A_{w}=A^{w}= A_{m_{1}} \ldots A_{m_{n}}$ is linear in 
the reduced momentum, while (\ref{ak}) and 
its hermitian conjugate  (\ref{al})  are  ``local but non-polynomial'' because
they are linear in the reduced momenta. 

Although we will not discuss it explicitly here, the phenomenon of 
this section also generates new large N field identifications (see 
Ref.~\cite{us}) in the unreduced theory.

\section{Dual Basis}%5

\subsection{Definition}%5.1

We wish to find new polynomials $\{T_{w}(\phi)\}$ which are vev-orthogonal
to the set  $\{G_{w}(\phi)\}$
\begin{equation}
<T_{w}(\phi)\; G_{\bar{w}^{\prime}}(\phi)> = \delta_{w,w^{\prime}}, 
\;\;\;\;\;\; T_{0}(\phi)=1 \label{am}
\end{equation}
and we will refer to the set of states $\{ <0| T_w(\phi) \}$ as the 
\emph{dual basis}.

Towards the construction of these polynomials, we first postulate a
generating function for the $T$'s
\bs \label{an}
\begin{equation}
Y = \frac{1}{1-\beta_{m} \phi_{m} + X(\beta)} = \sum_{w} 
\beta^{w} T_{w}(\phi) \label{cj}
\end{equation}
\begin{equation}
X(\beta) = \sum_{w} \; \beta^{w} \; X_{w}, \;\;\;\;\;\; 
X_{0} = 0
\end{equation}
\es
where $\beta_{m}$ is another free-algebraic source (like $\alpha_{m}$ above) and
the quantity $X(\beta)$ is to be determined. Note that the relations 
\begin{equation}
<T_{w}> = \delta_{w,0}, \;\;\;\;\;\; <Y> = 1
\end{equation}
follow from (\ref{am}) and (\ref{an}) respectively.

Next, follow the steps
\bs
\begin{eqnarray}
\la Y \tilde{B}^{\dagger}_{m} = \la [Y,-i \tilde{\pi}_{m}] = \la Y 
[1-\beta_{n} \phi_{n} + X, i \tilde{\pi}_{m}] Y \\ 
= \la Y \beta_{m} \ra \la Y = \beta_{m} \la Y
\end{eqnarray}
\es
where we have used  (\ref{k})  and (\ref{an}). Repeating this, we
 obtain
\begin{equation}
\la Y (\tilde{B}^{\dagger})^{w} \ra =\beta^{w} <Y> = \beta^{w}
\end{equation}
which, with (\ref{ag}), gives us the desired result (\ref{am}).

To compute $T_{w}$ and $X_{w}$ explicitly, multiply (\ref{cj}) on the 
left by the inverse of $Y$ to obtain
\begin{equation}
1 = \sum_{w}\beta^{w}T_{w} - \sum_{m,w}\beta^{mw}\phi_{m}T_{w} + 
\sum_{w,w^{\prime}} \beta^{ww^{\prime}}X_{w} T_{w^{\prime}}.
\end{equation}
Then, equating coefficients of each $\beta$ word, we find the recursion
relation for $T_{w}$
\begin{equation}
T_{mw} = \phi_{m}T_{w} - \sum_{w=w_{1}w_{2}} \; 
X_{mw_{1}}T_{w_{2}}, \;\;\;\;\;\; T_{0} = 1. \label{aq}
\end{equation}
Multiplying in the other order leads to
\begin{equation}
T_{wm} = T_{w} \phi_{m} - \sum_{w=w_{1}w_{2}} \; 
T_{w_{1}}X_{w_{2}m} \label{ar}
\end{equation}
and the vev's of these equations
\begin{equation}
X_{mw} = <\phi_{m}T_{w}> = <T_{w}\phi_{m}> = X_{wm} \label{as}
\end{equation}
determine the $X_{w}$'s and show that they have cyclic symmetry 
in the letters of their words.

\subsection{Examples}%5.2

Because the $T$'s and $X$'s are unfamiliar, we list the first few 
words of each:
\bs
\begin{equation}
T_{0}=1,\;\;\;\;\;\; T_{m} = \phi_{m} -X_{m}, \;\;\;\;\;\; T_{mn} = 
(\phi_{m}-X_{m})(\phi_{n}-X_{n}) - X_{mn}
\end{equation}
\begin{equation}
T_{mnp} = (\phi_{m}-X_{m})(\phi_{n}-X_{n})(\phi_{p}-X_{p}) - 
(\phi_{m}-X_{m})X_{np} - X_{mn}(\phi_{p}-X_{p}) - X_{mnp}
\end{equation}
\es

\bs
\begin{equation}
X_{0}=0, \;\;\;\;\;\; X_{m} = <\phi_{m}>, \;\;\;\;\;\; 
X_{mn} = <\phi_{m}\phi_{n}> - X_{m} X_{n}
\end{equation}
\begin{equation}
X_{mnp} = <\phi_{m}\phi_{n}\phi_{p}> - X_{m}X_{np} - 
X_{n}X_{mp}- X_{p}X_{mn} - 
X_{m}X_{n}X_{p}
\end{equation}
\begin{eqnarray}
X_{mnpq} &=& <\phi_{m}\phi_{n}\phi_{p}\phi_{q}> - X_{m}X_{npq} - 
X_{n}X_{mpq}- X_{p}X_{mnq} - 
X_{q}X_{mnp} \nonumber \\  &-& 
X_{mn}X_{pq}-X_{mq}X_{np} - 
X_{n}X_{m}X_{pq} -  X_{n}X_{p}X_{mq} - 
X_{n}X_{q}X_{mp} \nonumber \\  &-& X_{p}X_{m}X_{qn} - 
X_{q}X_{p}X_{mn} - X_{q}X_{m}X_{np} - 
X_{m}X_{n}X_{p}X_{q}.
\end{eqnarray}
\es
One  sees that the $X_{w}$'s so far match the planar connected parts discussed in 
Refs.~\cite{Brez, Cvit}, and one also 
sees that $T_{w}(\phi)$, with $<T_{w}>=\delta_{w,0}$, may be considered as a 
type of normal ordered 
product $:\phi^{w}:$ of the reduced fields.

\subsection{More general results}%5.3

From the recursive definitions (\ref{aq}-\ref{as}) we 
find 
\bs
\begin{equation}
T_{w}^{\;\;\;\dagger} = T_{\bar{w}}, \;\;\;\;\;\;X_{w}^{\;\;\;*} = 
X_{\bar{w}}
\end{equation}
\begin{equation}
<G_{\bar{w}}\; T_{w^{\prime}}> = \delta_{w,w^{\prime}}
\end{equation}
\es
as well as the following relations
\begin{equation}
T_{w} = \phi^{w} - 
\sum_{w=w_{1}w_{2}w_{3}}T_{w_{1}}X_{w_{2}}\phi^{w_{3}} \label{at} 
\end{equation}
\begin{equation}
T_{w}T_{w^{\prime}} = \sum_{w^{\prime\prime}}C_{w,w^{\prime},w^{\prime\prime}} 
T_{w^{\prime\prime}}, \;\;\;\;\;\;[w^{\prime\prime}]\leq [w]+[w^{\prime}] 
\label{au}
\end{equation}
\bs \label{av}
\begin{equation}
<T_{m}T_{w}> =X_{mw}(1-\delta_{w,0})
\end{equation}
\begin{equation}
<T_{mn}T_{w}> = X_{mnw}(1-\delta_{w,0}) + \sum_{\stackrel{w=w_{1}w_{2}}
{w_{1}, w_{2} \neq 0}} X_{nw_{1}}  \vspace{-12pt}  X_{mw_{2}}
\end{equation}
\hspace{2.5in} \vdots 
\es
\begin{equation}
Z(j) \equiv \sum_{w} < \phi^{w}> j^{w} = 1 + X(jZ(j)). \label{aw}
\end{equation}
In particular, (\ref{at})  can also be iterated to obtain the $T$'s. 
The relation in (\ref{au}) is an operator product expansion, whose sum 
obeys the selection rule shown because the $T$'s are finite polynomials 
in $\phi$. The list of  relations begun in (\ref{av}) shows correspondingly 
higher powers of $X_{w}$ when extended to more general words.

The final relation (\ref{aw}), with $j$ another free-algebraic source, is
proven in App.~D. This is  the standard relation~\cite{Brez, Cvit} between
the generating functions $Z$ and $X$ of planar and connected
planar correlators respectively, and completes the identification of $X_{w}$ as 
the planar connected part with $[w]$ legs.

\subsection{Completeness}%5.4

 The dual basis $\{<0|T_w(\phi)\}$ 
is complete because the $\{\phi^{w}\ra\}$ basis is complete, and 
therefore the basis $ \{ B^{\dagger \; w} \ra = G_w(\phi)|0>\}$ is also
complete~\footnote{A different argument for the completeness of 
$B^{\dagger w}\ra$ was given in Ref.~\cite{us}}.  This  gives the completeness
 statements
\begin{equation}
\textbf{1} = \sum_{w} G_{w}(\phi) \ra \la T_{\bar{w}}(\phi) = 
\sum_{w} T_{w}(\phi) \ra \la G_{\bar{w}}(\phi)
\end{equation}
and various consequences such as
\begin{equation}
\delta_{w, w^{\prime}} = \sum_{w^{\prime\prime}}<T_{\bar{w}}T_
{w^{\prime\prime}}> 
<G_{\bar{w}^{\prime\prime}}G_{w^{\prime}}>.
\end{equation}
Moreover, either set of polynomials $\{G_w(\phi)\}$ or 
$\{T_w(\phi)\}$ give
a complete basis~\footnote{There are questions which need further 
study concerning the domain of convergence of expansions in the $G_{w}$'s 
when infinite sums are involved. For example, in the case of one matrix with a 
pure $\phi^{4}$ action
 the functions $G_{w}(\phi)$ have  no linear term in 
 $\phi$ and yet Eq.~(\ref{bp}) says that an infinite sum of such 
 functions is equal to $\phi$.} for expansion of any polynomial in $\phi$
\begin{equation}
{\cal F}(\phi) = \sum_{w} G_{w}(\phi) <T_{\bar{w}}(\phi) {\cal F}(\phi)> =
 \sum_{w}
 T_{w}(\phi) <G_{\bar{w}}(\phi) {\cal F}(\phi)>. \label{ay}
\end{equation}
We have already encountered such an expansion in (\ref{au}).

Another operator product expansion which will be useful 
below  is
\begin{equation}
G_{w}G_{w^{\prime}} = \sum_{w^{\prime\prime}} G_{w^{\prime\prime}} 
<T_{\bar{w}^{\prime\prime}}G_{w}G_{w^{\prime}}>. \label{ax}
\end{equation}
The sum on the right of (\ref{ax}) is generally infinite, but finite in the case
of oscillators/free actions (see App.~B). It will also be useful to consider
 expansions of products of the master field:
 \bs
\begin{equation}
\phi_{m} = \sum_{w} X_{mw} \; G_{\bar{w}}(\phi) = \sum_{w} G_{w} (\phi) 
X_{\bar{w}m} \label{bp}
\end{equation}
\begin{equation}
\phi_{m} \phi_{n} = \sum_{w}(X_{mnw} + 
\sum_{w=w_{1}w_{2}}X_{nw_{1}}X_{mw_{2}})\;  \vspace{-12pt}   G_{\bar{w}}(\phi)
\end{equation}
\hspace{2.5in} \vdots \\ 
\es
The proof of these follow readily from  (\ref{ay}) and (\ref{av}).

\subsection{Operator form of the dual basis}%5.5

Recall that  $B^{\;\;\dagger}_{\bar{w}} \ra$ is the operator form of the basis
 $G_{w}(\phi) \ra$. To obtain the operator
form of the dual basis $\la T_w(\phi)$, we first define a new set of operators
$\bar{B}_{m}$
\begin{equation}
\bar{B}_{m} \equiv (E^{-1})_{mn}B_{n}. \label{br}
\end{equation}
The interacting Cuntz algebra (\ref{ba}) implies that these operators satisfy
a (dual basis) Cuntz algebra
\bs \label{bs}
\begin{equation}
\bar{B}_{m}B^{\dagger}_{n} = \delta_{m,n}, \;\;\;\;\;\; 
B^{\dagger}_{m}\bar{B}_{m} = 1 - \ra \la 
\end{equation}
\begin{equation}
\bar{B}_{m} \ra = \la B^{\dagger}_{m} = 0
\end{equation}
\es
although $\bar{B}_{m}$ and $B^{\dagger}_{m}$ are not hermitian conjugates. 
This curious fact will play a central role in the discussion of Sec.~7.

Next, we consider the product of any number of $\bar{B}$'s
\begin{equation}
 \bar{B}_{w} = \bar{B}^{w} = \bar{B}_{m_{1}} \ldots \bar{B}_{m_{n}}
\end{equation}
and note that
\begin{equation}
\la \bar{B}^{\bar{w^{\prime}}} G_{w} \ra = \la \bar{B}^{\bar{w^{\prime}}} 
B^{\dagger\;w} \ra = \delta_{w, w^{\prime}}.
\end{equation}
It follows that
\begin{equation}
\la (T_{\bar{w}^{\prime}} - \bar{B}_{\bar{w^{\prime}}}) G_{w} \ra = 0, \;\;\;\;\; 
\forall \; w
\end{equation}
and this gives the operator form of the dual basis
\bs
\begin{equation}
\la \bar{B}_{w} = \la T_{w}(\phi) 
\end{equation}
\begin{equation}
\textbf{1} = \sum_{w} \; B_{w}^{\;\;\dagger} \ra \la \bar{B}_{w}
\end{equation}
\es
because the basis $G_{w} \ra$ is complete.

The operator form of the dual basis gives us a number of new forms
for the planar connected parts
\begin{equation}
X_{\bar{w}mn} = <T_{\bar{w}m} \phi_{n}> = <\bar{B}_{\bar{w}m}\phi_{n}> = 
<\bar{B}_{\bar{w}} (E^{-1})_{mn}> = <T_{\bar{w}} (E^{-1})_{mn}> \label{ec}
\end{equation}
where we have used (\ref{ap}) and (\ref{ew}). Then the useful relation
\begin{equation}
(E^{-1}(\phi))_{mn} =  \sum_{w} X_{mn\bar{w}} G_{w}(\phi) \label{dd}
\end{equation}
follows immediately from (\ref{ay}).

For free random variables, we can say more.  Taken together, the form 
of $E$ in (\ref{fc}) and the final form of $X$ in (\ref{ec}) show that 
$X_{\bar{w}mn}
 \propto \delta_{m,n}$ in this case.  Then, the cyclic symmetry of $X_{w}$ tells 
us that the only nonzero planar connected parts are the ``single 
letter'' $X$'s
\begin{equation}
X_{w(m)} \equiv X_{m \ldots m} \neq 0, \;\;\;\;\;\; m=1\ldots d . \label{fb}
\end{equation}
This simple fact means that the computation of the planar connected 
parts (see Sec.~8) is one dimensional and, via Eq.~(\ref{aw}), the 
relation (\ref{fb}) explains many intricate identities among the planar parts.

\section{$BB^{\dagger}$}%6

We have so far verified all the relations of the  infinite
dimensional free algebra (\ref{c})
except for the $BB^{\dagger}$  relation (\ref{bc}). This relation requires a
combination of several of the principles we have discussed above, and will
be developed in stages.

Note first the relations
\bs
\begin{equation}
B_{wm}B^{\dagger}_{n} = B_{w}E_{mn}, 
\;\;\;\;\;\;B_{m}B_{wn}^{\;\;\;\;\;\;\dagger} = 
E_{mn}B_{w}^{\;\;\;\dagger}
\end{equation}
\begin{equation}
B_{wm}B_{w^{\prime}n}^{\;\;\;\;\;\;\dagger} = 
B_{w}E_{mn}B_{w^{\prime}}^{\;\;\;\dagger} \label{bb} 
\end{equation}
\es
which follow from (\ref{h}) alone. In (\ref{bb}), we see that this direction 
soon fails
to produce relations linear in $B_{w}$ and $B_{w}^{\;\;\dagger}$.

To obtain relations linear in $B$ and $B^{\dagger}$,  we consider the
product $B_{mw}B_{nw^{\prime}}^{\;\;\;\;\;\dagger}$  using the forms (\ref{bm}) 
of these
 operators in terms
of the interacting Cuntz operators.  Among the four resulting terms, the
only term quadratic in $B$, $B^{\dagger}$ is 
$B^{\dagger}_{m}G_{w}G_{w^{\prime}}B_{n}$. This
term may be ``linearized'' by first using the completeness relation 
(\ref{ax}) and then using the master constraints in the form 
(\ref{ac}). We find two alternative forms of the result:
\begin{eqnarray}
B_{mw} B_{nw^{\prime}}^{\;\;\;\;\;\dagger} = -[B^{\dagger}_{m} G_{w}
G_{\bar{w}^{\prime}n} + G_{mw} G_{\bar{w}^{\prime}} B_{n} - G_{mw} G_{\bar{w}^
{\prime}n}] 
\nonumber \\ 
+ \sum_{w^{\prime\prime}} <T_{\bar{w}^{\prime\prime}} G_{w} G_{\bar{w}^{\prime}}> 
[ B^{\dagger}_{m}G_{w^{\prime\prime}n} + G_{mw^{\prime\prime}}
 B_{n}  -G_{mw^{\prime\prime}n}] \label{bd} 
\end{eqnarray}
\begin{eqnarray}
B_{mw} B_{nw^{\prime}}^{\;\;\;\;\;\dagger} = [B_{mw} G_{\bar{w}^{\prime}n}+
G_{mw}B_{nw^{\prime}}^{\;\;\;\;\dagger} - G_{mw} G_{\bar{w}^{\prime}n}] 
\nonumber \\ 
- \sum_{w^{\prime\prime}} <T_{\bar{w}^{\prime\prime}} G_{w} G_{\bar{w}^{\prime}}> 
[ B_{mw^{\prime\prime}n} + 
B_{n\bar{w}^{\prime\prime}m}^{\;\;\;\;\;\;\;\;\dagger}
  -G_{mw^{\prime\prime}n}].  
\end{eqnarray}
These forms are linear in the  operators $B,B^{\dagger}$ but
 the coefficients are functions of $\phi$.  
 
A form
which is strictly linear in the generators $B_{w}, B_{w}^{\;\;\dagger}$ 
can be derived from (\ref{bd}) by again using the expansion (\ref{ax}) for the 
products of two $G$'s and then using the formulas (\ref{bm}) in reverse.  The 
result is
\bs
\begin{equation}
B_{mw}B_{nw^{\prime}}^{\;\;\;\;\;\;\dagger} = \sum_{w^{\prime\prime}} 
(B_{mw^{\prime\prime}}f_{w^{\prime\prime},w,n,w^{\prime}} + 
f^{*}_{w^{\prime\prime},w^{\prime},m,w}B_{nw^{\prime\prime}}^{\;\;\;\;\;\;
\dagger}) + E_{mw;nw^{\prime}}(\phi)
\end{equation}
\begin{equation}
f_{w^{\prime\prime},w,n,w^{\prime}}= <T_{\bar{w}^{\prime\prime}}G_{w}G_{\bar{w}^
{\prime}n}> - \sum_{u} \delta_{w^{\prime\prime},un} 
<T_{\bar{u}}G_{w}G_{\bar{w}^{\prime}}> \label{be}
\end{equation}
\begin{eqnarray}
E_{mw;nw^{\prime}} = G_{mw}G_{\bar{w}^{\prime}n} + 
\sum_{w^{\prime\prime}}[G_{mw^{\prime\prime}n}<T_{\bar{w}^{\prime\prime}}
G_{w}G_{w^{\prime}}> \nonumber \\ 
 - G_{mw^{\prime\prime}}<T_{\bar{w}^{\prime\prime}}G_{w}G_{\bar{w}^{\prime}n}>
 - G_{w^{\prime\prime}n}<T_{\bar{w}^{\prime\prime}}G_{mw}G_{\bar{w}^{\prime}}> ].
\end{eqnarray}
\es
One may compare this general structure with the simple oscillator 
results in  App.~B.

\section{Forms of the Master Field} %7

\subsection{Basic form}%7.1

 The form (\ref{bp}) of the master field in terms of the 
basis $G_{w}$
\begin{equation}
\phi_{m} = \sum_{w} X_{mw} G_{\bar{w}}(\phi) \label{bw}
\end{equation}
will be called the \emph{basic form} of the master field. All the other forms of
 the master field below follow from the basic form. 

\subsection{In terms of interacting Cuntz operators}%7.2

 The basis $G_{w}$ is a set of
 polynomials (see Sec.~3) 
 in $G_{m}$ and $E_{mn}$, which may in turn be written as
\begin{equation}
G_{m} = B_{m}+ B_{m}^{\dagger}, \;\;\;\;\;\; E_{mn} = B_{m}B_{n}^{\dagger}.
\end{equation}
These relations allow us to express the $G_{w}$'s and hence the 
master field (\ref{bw}) in terms of 
interacting Cuntz operators:
\begin{equation}
\phi_{m} = X_{m} + X_{mn}( B_{n}+ B_{n}^{\dagger}) + 
X_{mpn}(B_{n}B_{p} + B_{n}^{\dagger}( B_{p}+ B_{p}^{\dagger})) + \ldots \;.
 \label{eh}
\end{equation} 

\subsection{In terms of ÒordinaryÓ Cuntz operators}%7.3

 Recall the construction~\cite{us} of  ordinary 
Cuntz operators from the interacting Cuntz operators 
\bs
\begin{equation}
a_{m} = (E^{-\frac{1}{2}})_{mn}B_{n}, \;\;\;\;\;\; a_{m}^{\dagger} = 
B_{n}^{\dagger}(E^{-\frac{1}{2}})_{nm}
\end{equation}
\begin{equation}
a_{m}a_{n}^{\dagger} = \delta_{m,n}, \;\;\;\;\;\; a_{m}^{\dagger}a_{m} 
= 1- \ra \la , \;\;\;\;\;\; a_{m} \ra = \la a_{m}^{\dagger}=0
\end{equation}
\es
where $a^{\dagger}$ is the hermitian conjugate of $a$.
This allows us to express the master field (\ref{eh}) in terms of ordinary Cuntz 
operators:
\begin{equation}
\phi_{m} = X_{m} + X_{mn}((E^{\frac{1}{2}})_{nq}a_{q} + 
a^{\dagger}_{q}(E^{\frac{1}{2}})_{qn}) + \ldots \; .
\end{equation}

\subsection{Dual basis form}%7.4

 To express the master field in this form,  follow the steps
\begin{equation}
\bar{B}_{m} = (E^{-1})_{mn}B_{n} = \sum_{w} X_{mn\bar{w}} G_{w} (F_{n} 
+i \pi_{n}) = \sum_{w}X_{mn\bar{w}}(G_{wn} - B^{\dagger wn})
\end{equation}
where we have used the form (\ref{af}) for $B_{w}^{\;\;\dagger}$ and the 
identities (\ref{ad}),(\ref{dd}). Adding $X_{m}$, 
we obtain the dual basis form of the master field
\begin{equation}
\phi_{m} =\sum_{w}X_{mw}G_{\bar{w}} = \bar{B}_{m} + 
\sum_{w}X_{m\bar{w}} B^{\dagger w}. \label{bt}
\end{equation}
Recall from (\ref{bs}) that the operators $\bar{B}_{m}, B^{\dagger}_{m}$ 
 (with  $\bar{B}_{m}^{\dagger} \neq B_{m}^{\dagger}$)
also satisfy an ordinary   
Cuntz algebra. Other ways of writing the dual basis form include
\bs
\begin{equation}
\phi_{m} = \bar{B}_{m} + \sum_{w}X_{mw} B_{w}^{\;\;\;\dagger}
= \bar{B}_{m}(1+\bar{X}(B^{\dagger})) \label{de}
\end{equation}
\begin{equation}
\bar{X}(\beta) \equiv \sum_{w} \beta^{w}X_{\bar{w}}, \;\;\;\;\;\; 
\la \bar{X}(B^{\dagger}) = 0.
\end{equation}
\es
Here, the first  form in (\ref{de}) emphasizes that the master field is linear 
in the generators  of the infinite dimensional free algebra, 
and $\bar{X}(B^{\dagger})$ in the second form is an alternate
 generating function of the planar 
connected parts.
 
 Note that the forms (\ref{bt}) and (\ref{de}) of the master field (and other 
 forms throughout 
 this section which are equal to $\phi_{m}$ in (\ref{bw})) appear to involve 
 the reduced momenta $\pi_{m}$ in the  creation and 
 annihilation operators.  However, 
 as the reader is encouraged to verify, all such $\pi$ terms cancel.
 
 \subsection{Second dual basis form}%7.5
 
  In spite of appearances, the dual basis
 form (\ref{de}) of the 
master field is hermitian (as are all the previous forms), which tells us that
\begin{equation}
\phi_{m} = \phi_{m}^{\dagger} = \bar{B}^{\dagger}_{m} + \sum_{w} 
X_{m\bar{w}} B^{w}, 
\;\;\;\;\;\;\bar{B}^{\dagger}_{m} = 
B^{\dagger}_{n}(E^{-1})_{nm}.\label{df}
\end{equation}
The operators $B, \bar{B}^{\dagger}$, with $B^{\dagger} \neq \bar{B}^{\dagger}$,
 form another (second dual basis) Cuntz algebra
\bs
\begin{equation}
B_{m}\bar{B}^{\dagger}_{n} = \delta_{m,n}, \;\;\;\;\;\; 
\bar{B}^{\dagger}_{m}B_{m} = 1-\ra \la , \;\;\;\;\;\;
B_{m}\ra = \la \bar{B}^{\dagger}_{m} = 0 
\end{equation}
\begin{equation}
\bar{B}_{w}^{\;\;\dagger} \ra = T_{\bar{w}}(\phi) \ra
\end{equation}
\begin{equation}
\textbf{1} = \sum_{w}\; \bar{B}_{w}^{\;\;\dagger} \ra \la B_{w}
\end{equation}
\es
and we see that $\bar{B}_{w}^{\;\;\dagger}$ creates the ket form of the dual 
basis. 

\subsection{Non-hermitian forms}%7.6

 Because the two sets of operators 
($a_{m},a^{\dagger}_{n}$) and ($\bar{B}_{m}, B^{\dagger}_{n}$) both satisfy the
 Cuntz algebra,  the two sets are related by a similarity transformation $S$
\bs \label{bu}
\begin{equation}
S a_{m}S^{-1} = \bar{B}_{m} = (E^{-1}(\phi))_{mn}B_{n} = 
(E^{-\frac{1}{2}}(\phi))_{mn} a_{n}
\end{equation}
\begin{equation}
S a^{\dagger}_{m}S^{-1} = B_{m}^{\dagger} = 
a_{n}^{\dagger}(E^{\frac{1}{2}}(\phi))_{nm}
\end{equation}
\es
and $S$ cannot be unitary because $B^{\dagger}$ is not the hermitian conjugate
 of $\bar{B}$. 

Then we see
that the dual form  of the master field in (\ref{de}) is the hermitian 
counterpart  of the non-hermitian Gopakumar-Gross form $M_{m}$ of the master 
field:
\bs \label{bv}
\begin{equation}
\phi_{m} = S M_{m}S^{-1}
\end{equation}
\begin{equation}
M_{m} = a_{m} + \sum_{w} X_{m\bar{w}} a^{\dagger w} = 
a_{m}(1+\bar{X}(a^{\dagger}))
\neq M_{m}^{\dagger}. \label{dg}
\end{equation}
\es
Our algebraic derivation of (\ref{dg}) complements the diagrammatic 
derivation~\footnote{Unfortunately, Gopakumar and Gross give the 
similar but incorrect result $M_{m} = a_{m}+ \sum_{w}X_{mw}a^{\dagger 
\; w}$, as we ourselves did in an earlier version.  To check that 
(\ref{dg}) is 
in fact the correct form, evaluate $<M_{m}M_{n}M_{p}>$ using 
the Cuntz algebra for $a,a^{\dagger}$.} in Ref.~\cite{Gopa}.  The one-matrix form 
$M=a+\Sigma_{m}c_{m+1}a^{\dagger m}$ of the non-hermitian master field
was determined earlier in Ref.~\cite{Voic}.

The hermitian conjugate of the Gopakumar-Gross form, which also serves as a 
master field,  is related to the second dual form (\ref{df}) of the hermitian 
master field as follows
\bs
\begin{eqnarray}
S^{-1\; \dagger}a_{m} S^{\dagger} = B_{m} = (E^{\frac{1}{2}})_{mn}a_{n} \\
S^{-1\; \dagger}a^{\dagger}_{m} S^{\dagger} = \bar{B}_{m}^{\dagger} = 
B_{n}^{\dagger}(E^{-1})_{nm} = a_{n}^{\dagger}(E^{-\frac{1}{2}})_{nm} \\
\phi_{m} = \phi_{m}^{\dagger}= S^{-1\; \dagger}M_{m}^{\dagger} S^{\dagger} \\ 
M_{m}^{\dagger} = a_{m}^{\dagger} + \sum_{w}X_{m\bar{w}} a^{w} = 
(1+\bar{X}(a)) a_{m}^{\dagger}.
\end{eqnarray}
\es
These relations are nothing but the hermitian conjugate of 
(\ref{bu}) and (\ref{bv}).

\subsection{In terms of planar correlators}%7.7

  The relation 
(\ref{ff}) can be read as
\bs \label{dw}
\begin{equation}
\bar{Z}(j) = 1+\bar{X}(B^{\dagger}), \;\;\;\;\;\; \bar{Z}(j) = \sum_{w} 
j^{\bar{w}} <\phi^{w}> 
\end{equation}
\begin{equation}
B^{\dagger}_{m} = j_{m} \bar{Z}(j), \;\;\;\;\;\; j_{m} = B^{\dagger}_{m} 
\bar{Z}^{-1}(j) \label{dv}
\end{equation}
\begin{equation}
\bar{B}_{m} j_{n} = \delta_{m,n} \bar{Z}^{-1}(j), \;\;\;\;\;\; 
\la \bar{Z}(j) = \la 
\end{equation}
\es
where $\bar{Z}(j)$ is  an alternate generating function for
 planar correlators.  The 
``quantum source'' $j_{m}$ lives in a fourth Cuntz algebra
\bs
\begin{equation}
\frac{\partial}{\partial j_{m}} \equiv \bar{Z}(j) \bar{B}_{m}, \;\;\;\;\;\;
\bar{B}_{m} = \bar{Z}^{-1}(j) \frac{\partial}{\partial j_{m}} \label{ea}
\end{equation}
\begin{equation}
\frac{\partial}{\partial j_{m}} j_{n} = \delta_{m,n}, \;\;\;\;\;\;
j_{m}\frac{\partial}{\partial j_{m}} = 1-\ra \la \label{ej}
\end{equation}
\begin{equation}
\frac{\partial}{\partial j_{m}} \ra = \la j_{m} = 0
\end{equation}
\es
which follows from (\ref{dw}) and the Cuntz algebra (\ref{bs}) of $\bar{B}$ and 
$B^{\dagger}$.

This gives the forms of the master field
\bs
\begin{equation}
\phi_{m} = \bar{B}_{m} \bar{Z}(j) = \bar{Z}^{-1}(j) \frac{\partial}
{\partial j_{m}}\bar{Z}(j)  \label{du}
\end{equation}
\begin{equation}
\phi^{w} =  \bar{Z}^{-1}(j) (\frac{\partial}{\partial j})^{w}\bar{Z}(j)  \label{eb}
\end{equation}
\es
in terms of the planar correlators. 

\subsection{In terms of planar 1PI parts}%7.8

  The master field 
can also be written as a function of the planar 
connected one particle irreducible (1PI) parts.  To see this, we 
first decompose the dual basis form of the master field (\ref{de}) into its 
classical part 
$\Phi_{m}$ and its quantum part $\bar{B}_{m}$
\bs
\begin{equation}
\phi_{m} = \Phi_{m} + \bar{B}_{m}
\end{equation}
\begin{equation}
\Phi_{m}(B^{\dagger}) \equiv \bar{B}_{m} \bar{X}(B^{\dagger}) = \sum_{w}
 X_{m\bar{w}} B^{\dagger \; w}
\end{equation}
\begin{equation}
\Phi_{m} B^{\dagger}_{m} = B^{\dagger}_{m} \Phi_{m} = \bar{X}(B^{\dagger}).\label{dm}
\end{equation}
\es
Our definition of the classical part $\Phi_{m}$ agrees with  
the field called $\Phi$ in Ref.~\cite{Cvit}, but the identities in (\ref{dm}) are 
new. 

The planar effective action $\Gamma(\Phi)$ is  defined as
\bs \label{dq}
\begin{equation}
\Gamma(\Phi) \equiv \Phi_{m} B^{\dagger}_{m} = B^{\dagger}_{m} \Phi_{m} = 
\bar{X}(B^{\dagger}) = \sum_{w} \Gamma_{w} \Phi^{w} \label{ei}
\end{equation}
\begin{equation}
\bar{B}_{m} \Gamma(\Phi) = \Phi_{m}, \;\;\;\;\;\; \la \Gamma(\Phi) = 0
\end{equation}
\es
where $\Gamma_{w}$ is the cyclically symmetric planar 1PI part with $[w]$ legs. 
 This 
definition of $\Gamma(\Phi)$ follows Ref.~\cite{Brez} but differs by a 
minus sign from the definition of Ref.~\cite{Cvit}, and we note in particular
 that the Legendre transform defined in Ref.~\cite{Cvit}
\begin{equation}
\bar{X}(B^{\dagger}) = - \Gamma(\Phi) + B^{\dagger}_{m}\Phi_{m} + 
\Phi_{m}B^{\dagger}_{m}
\end{equation}
is satisfied trivially by (\ref{ei}).

  Then the master field 
can be written as
\begin{equation}
\phi_{m} = \Phi_{m} + \bar{B}_{m} = \bar{B}_{m}(1+\Gamma(\Phi)) \label{dr}
\end{equation}
by changing variables from $B^{\dagger}$ to $\Phi$. But this is only 
half the job because we also want to find the Cuntz algebra in which 
$\Phi_{m}$ resides.

This is most easily done in the case $X_{m}=0$ (no tadpoles), which we 
assume below.  In this case, one has the additional relations
\bs
\begin{equation}
\Phi_{m} = B^{\dagger}_{n} \gamma_{nm} = \gamma_{mn}B^{\dagger}_{n}
\end{equation}
\begin{equation}
\gamma_{mn}(B^{\dagger}) = \sum_{w} X_{mn\bar{w}} B^{\dagger \; w}, 
\;\;\;\;\;\; \gamma_{mn} \ra = (E^{-1})_{mn} \ra
\end{equation}
\begin{equation}
\bar{B}_{m}\Phi_{n} = \gamma_{mn}
\end{equation}
\es
and $\gamma_{mn}$ is invertible because it begins with $X_{mn}$.  This 
gives us the Cuntz algebra of $\Phi_{m}$:
\bs
\begin{equation}
\frac{\partial}{\partial \Phi_{m}} \equiv (\gamma^{-1})_{mn}\bar{B}_{n} \label{ek}
\end{equation}
\begin{equation}
\frac{\partial}{\partial \Phi_{m}} \Phi_{n} = \delta_{m,n}, 
\;\;\;\;\;\; \Phi_{m} \frac{\partial}{\partial \Phi_{m}} = 1-\ra \la 
\end{equation}
\begin{equation}
\frac{\partial}{\partial \Phi_{m}} \ra = \la \Phi_{m} = 0
\end{equation}
\es
and we may now express the dual basis Cuntz operators as
\bs
\begin{equation}
B^{\dagger}_{m} = \Phi_{n}(\gamma^{-1})_{nm} = (\gamma^{-1})_{mn} \Phi_{n}
\label{el}
\end{equation}
\begin{equation}
\bar{B}_{m} = \gamma_{mn} \frac{\partial}{\partial \Phi_{n}}. \label{em}
\end{equation}
\es
Moreover, the relation
\begin{equation}
B^{\dagger}_{m} = \frac{\partial}{\partial \Phi_{m}} \Gamma(\Phi) \label{en}
\end{equation}
now follows from (\ref{ek}), (\ref{ej}) and (\ref{el}).

Our next task is to find the $\Phi$ dependence of 
$\gamma_{mn}(B^{\dagger})$.  Note first that
\begin{equation}
\Gamma(\Phi) = \Phi_{m} \Phi_{n} (\gamma^{-1})_{nm}
\end{equation}
follows from (\ref{ei}) and (\ref{el}), and this gives us the desired result
\begin{equation}
(\gamma^{-1}(\Phi))_{mn} = \frac{\partial}{\partial \Phi_{m}} 
\frac{\partial}{\partial \Phi_{n}} \Gamma(\Phi).
\end{equation}
Using (\ref{em}) and (\ref{en}) in (\ref{dr}), we have found the forms of the
 master field
\begin{equation}
\phi_{m} = \gamma_{mn}(\Phi) \frac{\partial}{\partial \Phi_{n}}
(1+\Gamma(\Phi)) = \Phi_{m} + \gamma_{mn}(\Phi) \frac{\partial}{\partial 
\Phi_{n}} \label{eu}
\end{equation}
in the $\Phi, \frac{\partial}{\partial \Phi_{}}$ basis.

Comparing these two forms of the master field (or the two forms of 
$B^{\dagger}$ in (\ref{el})), we also find the consistency relation
\begin{equation}
\frac{\partial}{\partial \Phi_{m}} \Gamma(\Phi) = 
(\frac{\partial}{\partial \Phi_{m}}\frac{\partial}{\partial 
\Phi_{n}} \Gamma(\Phi)) \Phi_{n}
\end{equation}
but this is only the statement that $\Gamma_{w}$ is cyclically 
symmetric.

\section{Forms of the Schwinger-Dyson Equations}%8

In this section, we use the forms of the master field to quickly 
derive a number of new free-algebraic forms of the large N Schwinger-Dyson 
equations for action theories.\footnote{Another form of the 
Schwinger-Dyson equations follows as null state Ward identities of 
the infinite dimensional free algebra (see App.~E).}  The first form in Subsec.~8.1 is 
novel, and the rest, although packaged differently,  are closely related to 
known free-algebraic formulations~\cite{Brez, Cvit, Doug1, Doug2}.  In all our 
formulations, the dynamical input is stored in the  
operators  $G_{m}(\phi), E_{mn}(\phi)$ of the interacting Cuntz 
algebra (\ref{o}),(\ref{ba}).

\subsection{The basic form as a computational system}%8.1

We consider first the  basic form of the master field 
\begin{equation}
\phi_{m} = \sum_{w} X_{mw} G_{\bar{w}}(\phi) \label{bw1}
\end{equation}
which, by matching $\phi$ dependence on left and right, is itself a computational 
system for the planar connected parts. 

We illustrate this by studying the classical limit of the system. Reinstating 
$\hbar$ temporarily, we find that 
\begin{equation}
 G_{m} = O(\hbar^{0}), \;\;\;\;\;\; E_{mn}=O(\hbar)
\end{equation}
because $E^{mn}$ in (\ref{dp}) is a commutator.
The classical limit of (\ref{bw1})
\begin{equation}
\phi_{m} \simeq \sum_{w} X_{mw} G^{\bar{w}}, \;\;\;\;\;\; 
G_{w} \simeq G^{w} = G_{m_{1}} \ldots G_{m_{n}} \label{eo}
\end{equation}
is then obtained by neglecting all $E$ terms in the $G_{w}$'s (see Eq.(\ref{bx})). 

For definiteness, we consider the solution of this equation for the general 
quartic interaction
\begin{equation}
G_{m} = 2 \omega_{m}\phi_{m} + \lambda_{mnpq}\phi_{n}\phi_{p}\phi_{q} \label{ca}
\end{equation}
where $\lambda_{mnpq}$ is cyclically symmetric in its indices.
In this case, (\ref{eo}) contains only odd powers of $\phi$ and we may set the 
coefficients of each odd power to zero,  obtaining the list of 
equations
\bs
\begin{equation}
\phi:\;\;\;\;\; \phi_{m} = X_{mn} 2 \omega_{n} \phi_{n}
\end{equation}
\begin{equation}
\phi^{3}: \;\;\;\;\;  0 = X_{mn} \lambda_{npqr}\phi_{p}\phi_{q}\phi_{r} + 
X_{mnpq} 2\omega_{q} \phi_{q} 2\omega_{p}\phi_{p} 2  \vspace{-12pt} 
\omega_{n}\phi_{n}
\end{equation}
\hspace{2.5in} \vdots 
\es \\
The master field $\phi_{m}$ is a free variable (with no relations), so the unique 
solution of this list  is easily obtained: 
\bs \label{cb}
\begin{equation}
X_{mn} = \frac{1}{2\omega_{m}} \delta_{m,n}, \;\;\;\;\;\;
X_{mnpq} = - \frac{\lambda_{mqpn}}{2\omega_{m}2\omega_{n}2\omega_{p}
2\omega_{q}}
\end{equation}
\begin{equation}
X_{mnpqrs} = \frac{1}{\prod 2\omega} \sum_{t} \frac{1}{2\omega_{t}}
(\lambda_{msrt}\lambda_{tqpn} + \lambda_{nmst} \lambda_{trqp}  \vspace{-6pt}
+ \lambda_{pnmt}  \vspace{-8pt} \lambda_{tsrq})
\end{equation}
\hspace{2.5in} \vdots 
\es \\
These results are recognized as the tree-graph contributions to the planar 
connected parts.

For the special case of free random variables, the basic form 
(\ref{bw1}) decouples into d one-matrix problems with $\bar{X}=X$
\begin{equation}
\phi_{m} = \sum_{w(m)} X_{w(m)} G_{w(m)} (\phi_{m})
\end{equation}
according to Eqs.~(\ref{fc}), (\ref{fb}) and (\ref{bx}).  The one-matrix bases 
$G_{w(m)}(\phi_{m})$ have the decoupled form discussed in Subsec.~3.3.

Other relations of this type, e.g. Eq.~(\ref{dd}), may also be considered as 
computational systems.

\subsection{Dual basis system}%8.2

The planar connected parts $\bar{X}(B^{\dagger})$ satisfy
\bs \label{ey}
\begin{equation}
B^{\dagger}_{m} + E_{mn}(\phi)\bar{B}_{n} = G_{m}(\phi) \label{bz}
\end{equation}
\begin{equation} 
\phi_{p} = \bar{B}_{p}(1+\bar{X}(B^{\dagger})) \label{bz2}
\end{equation}
\es
which we record together as the \emph{dual basis system} 
\begin{equation}
B^{\dagger}_{m} + E_{mn}(\bar{B}(1+\bar{X}))\bar{B}_{n} = 
G_{m}(\bar{B}(1+\bar{X})). \label{bz3}
\end{equation}
To derive this system, start with $G_{m} = B_{m} + B^{\dagger}_{m}$,  go to the 
dual 
basis with (\ref{br}) and use the dual basis form (\ref{bz2}) of the master field.

 We have checked that the system (\ref{bz3}), 
 although packaged differently,  is equivalent to the Schwinger-Dyson 
equations derived diagrammatically for the planar connected parts 
in Ref.~\cite{Cvit}. 
In particular, our Cuntz operators $\bar{B}_{m}$ act on $\bar{X}(B^{\dagger})$ as
the  operator $\frac{\delta}{\delta J_{m}}$ of~\cite{Cvit} acts on their
$W(J)$, but the two operators are not the same because
\begin{equation}
[\bar{B}_{m},c] = 0, \;\;\;\;\;\; \frac{\delta}{\delta J_{m}} c = 0
\end{equation}
for any c-number c.  The $E$ term in (\ref{bz3}) collects the results of this 
difference.
 In what follows, we make some additional remarks on the structure of the 
dual basis system.
  
We begin by discussing this system in the case of one matrix, where right 
multiplication by powers of $B^{\dagger}$ gives the simple equation~\footnote
{This equation gives the large $\beta$ 
form $X(\beta) \sim c^{-\frac{1}{p}}\beta^{1+\frac{2}{p}}$ when $G(\phi) \sim 
c \phi^{p}$ at large $\phi$.} 
\begin{equation}
E(\frac{\psi}{\beta}) -G(\frac{\psi}{\beta})\beta + \beta^{2} = 0, 
\;\;\;\;\;\;\psi(\beta)=1+X(\beta), \;\;\;\;\;\;\psi(0)=1 \label{ds}
\end{equation}
for any interaction. (We have replaced $B^{\dagger}$ by a commuting source 
$\beta$.) In the special case of the quartic interaction (see 
(\ref{et})), this reads
\begin{equation}
\lambda \psi^{2}(\psi-1) + \beta^{2}(m^{2}(\psi-1) - \lambda X_{2}
-\beta^{2}) = 0
\end{equation}
and, except that $X_{2}$ appears as an unknown,  this  is the cubic 
equation found  in Ref.~\cite{Brez} for this interaction. In fact, the equation
 determines $X_{2}$ along with the rest of $X(\beta)$ in a perturbative 
  or semiclassical expansion. To begin the perturbation theory, set 
  $\lambda=0$ to find $\psi(\beta) = 1+\frac{\beta^{2}}{m^{2}}$. More 
  general perturbation theory is discussed in App.~F.
  
For the special case of free random variables, the dual basis 
system (\ref{bz3}) decouples into d one-matrix systems
\bs
\begin{eqnarray}
&&B^{\dagger}_{m} + E_{m}(\phi_{m})\bar{B}_{m} = G_{m}(\phi_{m}) \\ 
&&\phi_{m} = \bar{B}_{m} + \sum_{w(m)}X_{mw(m)}B^{\dagger\;w(m)} 
\end{eqnarray}
\es
which comprise d decoupled systems of the form (\ref{ds}).

The classical limit of the full system  (\ref{bz3}) is
\begin{equation}
B^{\dagger}_{m} \simeq G_{m}(\phi), \;\;\;\;\;\;\phi_{p} \simeq 
\bar{B}_{p}\bar{X}(B^{\dagger}) = \sum_{w}X_{p\bar{w}}B^{\dagger\; w} \label{ev}
\end{equation}
because $(1+\bar{X})$ in (\ref{bz2}) should be replaced by the dimensionless 
combination $(1+\bar{X}/\hbar)$.  As an example, the classical limit (\ref{ev}) reads
\begin{eqnarray}
0 =\!\!\!\!\!\!\!\! &&(B^{\dagger}_{m} - 2\omega_{m}X_{mn}B^{\dagger}_{n}) - 
\nonumber \\ 
&&(2\omega_{m}X_{mnpq}B^{\dagger}_{q}B^{\dagger}_{p}B^{\dagger}_{n}
+ \lambda_{mnpq}X_{nr}B^{\dagger}_{r}X_{ps}B^{\dagger}_{s}
X_{qt}B^{\dagger}_{t}) + \ldots
\end{eqnarray}
for the general quartic interaction (\ref{ca}). Setting each power of 
$B^{\dagger}$ to zero separately, we
find the same tree graphs (\ref{cb}) for the planar connected parts.

An equivalent form of the dual basis system (\ref{bz3}) is 
\begin{equation}
a_{m}^{\dagger} + E_{mn}(M) a_{n} = G_{m}(M), \;\;\;\;\;\;
M_{p}= a_{p}(1+\bar{X}(a^{\dagger})) \label{cc}
\end{equation}
in terms of ordinary Cuntz operators and the Gopakumar-Gross form of 
the master field.

The other forms of the planar Schwinger-Dyson equations below are 
the forms taken by Eq.~(\ref{bz3}) in different bases.

\subsection{Equation for the planar correlators}%8.3

The generating function $\bar{Z}(j)$ of planar correlators satisfies
\begin{equation}
j_{m}\bar{Z}(j) - G_{m}(\bar{Z}^{-1}(j)\frac{\partial}{\partial j}
\bar{Z}(j)) + E_{mn}
(\bar{Z}^{-1}(j)\frac{\partial}{\partial j}\bar{Z}(j))\bar{Z}^{-1}(j)
\frac{\partial}{\partial 
j_{n}} = 0.
\end{equation}
To derive this, use (\ref{bz}), (\ref{dw}), (\ref{ea}) and the form 
(\ref{du}) of the master 
field.  This can be simplified to
\begin{equation}
(\bar{Z}(j) j_{m} - G_{m}(\frac{\partial}{\partial j}))\bar{Z}(j)
 + E_{mn}(\frac{\partial}{\partial j})\frac{\partial}{\partial j_{n}}=0
  \label{ep}
\end{equation}
for any polynomial interaction.

  For the one-matrix case ($\bar{Z}=Z$) the system (\ref{ep}) reduces  to the 
quadra- \\ tic equation
\begin{equation}
(jZ(j))^{2} - G(\frac{1}{j}) jZ(j) + E(\frac{1}{j}) = 0, \;\;\;\;\;\;
 Z(0)=1 \label{er}
\end{equation}
for any interaction.  This equation may also be obtained from Eq. 
(\ref{ds}) and 
\begin{equation}
\psi(B^{\dagger}) = Z(j), \;\;\;\;\;\; B^{\dagger} = jZ(j), 
\;\;\;\;\;\; \frac{\psi(B^{\dagger})}{B^{\dagger}} = \frac{1}{j} 
\end{equation}
which is the one dimensional form of $\psi=1+X$ and (\ref{dw}).

Again, the relations (\ref{ep}) or (\ref{er}) are equivalent to those given in 
Ref.~\cite{Cvit}, 
although ours are packaged differently.  In particular, our ``derivative'' 
with respect to $j$ is a Cuntz operator satisfying 
\begin{equation}
\frac{\partial}{\partial j_{m}} c = 
c \frac{\partial}{\partial j_{m}}
\end{equation}
 when $c$ is a c-number, and 
\emph{not} the rule $\frac{\delta c}{\delta j_{m}} = 0$ satisfied 
by the derivative in Ref.~\cite{Cvit}.  The difference between these two 
operators is again collected in the $E$ term of (\ref{ep}).

\subsection{Equation for the planar effective action}%8.4

The planar effective action $\Gamma(\Phi)$ satisfies 
\begin{equation}
\frac{\partial}{\partial \Phi_{m}} \Gamma(\Phi) + E_{mn}(\Phi+\gamma 
\frac{\partial}{\partial \Phi}) \gamma_{np}(\Phi) \frac{\partial}{\partial 
\Phi_{p}} = G_{m}(\Phi+\gamma\frac{\partial}{\partial \Phi}). \label{eq}
\end{equation}
To derive this system, use (\ref{em}), (\ref{en}) and (\ref{eu}) in 
(\ref{bz}).
Although packaged differently, this system is equivalent to the 
equation for $\Gamma$ given in Ref.~\cite{Cvit}. (Again, our Cuntz operator 
$\frac{\partial}{\partial \Phi}$ commutes with c-numbers and so is 
not equal to the operator $\frac{\delta}{\delta \Phi}$ of~\cite{Cvit}.)

For the classical limit of (\ref{eq}), we know to neglect $E$ and the quantum 
part $\bar{B} = \gamma \frac{\partial}{\partial \Phi}$ of the master field.
  This gives immediately the
classical limit of the planar effective action
\begin{equation}
\Gamma(\Phi) \simeq \Phi_{m} G_{m} (\Phi)
\end{equation}
for any theory.

For the general one-matrix model, the system (\ref{eq}) simplifies to
\begin{equation}
\{\Gamma(\Phi) - \Phi G(\Phi[1+\Gamma^{-1}(\Phi)])\}\Gamma(\Phi) + \Phi^{2}
E(\Phi[1+\Gamma^{-1}(\Phi)]) = 0.
\end{equation}
This equation can also be obtained directly from (\ref{ds}) by the 
transformation
\bs \label{dt}
\begin{equation}
\psi = 1+X(B^{\dagger}) = 1+\Gamma(\Phi)
\end{equation}
\begin{equation}
B^{\dagger} = \beta = \Gamma(\Phi)/\Phi 
\end{equation}
\es
which is the one-dimensional form of (\ref{dq}). The relations 
(\ref{dt}) were pointed 
out in~\cite{Brez}, and we have checked for the quartic case  (\ref{et}) that 
the resulting cubic equation is in agreement with that given there.

\vspace{1.2cm}

\noindent {\bf ACKNOWLEDGEMENTS}

For helpful discussions, we thank J.~de~Boer, J.~Evslin, H.~Ooguri, 
\\ 
C.~Schweigert and J.~Wang.

     The work of  M.~B.~H. was supported in part by the Director, Office of 
Energy Research, Office of Basic Energy Sciences, of the U.S. 
 Department 
of Energy under Contract DE-AC03-76F00098 and in part by the National 
Science Foundation under grant PHY95-14797.

\vskip 1.0cm
\setcounter{equation}{0}
\def\theequation{A.\arabic{equation}}
\boldmath
\noindent{\bf Appendix A. Large N as Higher-Dimensional Classical Solution}%A
\unboldmath
\vskip 0.5cm

The fifth-time formulation~\cite{Green} of any Euclidean action theory allows us 
to compute the large N limit of the action theory as a classical solution of 
a higher-dimensional theory, in parallel with the AdS/CFT 
correspondence [3-5]. There is great latitude in the choice 
of the fifth-time theory, but  any choice will give the same large N averages 
for the original theory. Moreover, other methods of 
higher-dimensional extension are known (see e.g. Ref.~\cite{Pari}) and others 
still can be invented.

As an illustration,  we consider the action theory 
\begin{equation}
S = Tr(\frac{1}{2} m^{2}\phi^{2} + \frac{\lambda}{4N} \phi^{4})
\end{equation}
and we will choose the higher-dimensional extension (overdot is fifth-time 
derivative)
\bs
\begin{equation}
H_{5} = \frac{1}{2} Tr(B^{\dagger}B) = \frac{1}{2} Tr(\pi^{2}) + V_{5}
\end{equation}
\begin{equation}
V_{5} = \frac{1}{8} Tr[(m^{2}\phi + \frac{\lambda}{N} \phi^{3})^{2}] -
\frac{1}{4} [m^{2}N^{2} + 2\lambda Tr(\phi^{2}) + \frac{\lambda}{N} 
(Tr \phi)^{2}]
\end{equation}
\begin{equation}
S_{5} = \int dt (Tr(\frac{1}{2} \dot{\phi}^{2}) - V_{5})
\end{equation}
\es
which is a special case of the simple $H_{5}$ in Eq. (\ref{e}). 

Now we may follow Ref.~\cite{usold} to consider the phase-space master field, 
which solves the higher-dimensional classical equations of motion. Using 
Appendices C and E of Ref.~\cite{us} and in particular Eq.~(\ref{ce}) of the 
present paper,  we find the reduced classical equations of motion   
\bs
\begin{equation}
\dot{\phi} = \pi, \;\;\;\;\;\; \dot{\pi} = - V^{\prime}
\end{equation}
\begin{equation}
V =\frac{1}{8} (s^{\prime})^{2} - \frac{\lambda}{2}\phi(\phi + <\phi>), 
\;\;\;\;\;\; s \equiv \frac{m^{2}}{2}\phi^{2} + 
\frac{\lambda}{4}\phi^{4} \label{cf}
\end{equation}
\es
and the ground state density
\begin{equation}
\rho(\phi) = \frac{1}{\pi} {\sqrt {2(\epsilon - V(\phi))}}, 
\;\;\;\;\;\; \int d\phi \rho(\phi) = 1 \label{cg}
\end{equation}
from which the original action averages can be computed. (One may 
set  $<\phi>=0$ by symmetry.)

We note that, relative to the discussion of Ref.~\cite{Brez},
 the higher-dimension-\newline al extension has done the relevant Hilbert 
inversion for us
\begin{equation}
\frac{1}{2} s^{\prime}(\phi) = F(\phi) = \int dq \; \frac{\cal P}{\phi-q} 
\rho(q) \label{ch}
\end{equation}
(F is given in (\ref{by})) and moreover the extension has given us the ground 
state density $\rho$ in the higher-dimensional form (\ref{cg}). 
Using (\ref{e}), these features persist for the higher-dimensional solution 
of any one-matrix action theory. 

Finally, Eqs.~(\ref{v}), (\ref{cf}) and (\ref{cg}) tell us that
\begin{equation}
E = F^{2} + \pi^{2} \rho^{2} = 2\epsilon + \lambda \phi^{2}
\end{equation}
and we obtain
\begin{equation}
<\phi^{2}> = \frac{2\epsilon -m^{2}}{\lambda}
\end{equation}
on comparison with the form of $E$ in (\ref{by}).

\newpage

\vskip 1.0cm
\setcounter{equation}{0}
\def\theequation{B.\arabic{equation}}
\boldmath
\noindent{\bf Appendix B. Oscillators/Free Actions}%B
\unboldmath
\vskip 0.5cm

A number of simplifications occur for oscillator Hamiltonians and/or free action 
theories, which we treat together here in the oscillator notation (for 
free action theories,  $S=\frac{1}{2}\Sigma_{n}m^{2}_{n}Tr(\phi^{n}\phi^{n})$, 
replace $2\omega_{n}$ by $m^{2}_{n}$)
\begin{equation}
G_{m} = 2\omega_{m}\phi_{m}, \;\;\;\;\;\; E_{mn} = 2 
\omega_{m}\delta_{m,n}, \;\;\;\;\;\; X_{mn} = \frac{1}{2\omega_{m}}\delta_{m,n}.
\label{cs}
\end{equation} 
All other planar connected parts are zero.

Comparing the generating functions (\ref{s}) and (\ref{cj}), we find that the 
basis 
polynomials $G_{w}$ and the dual basis polynomials $T_{w}$ are proportional
\begin{equation}
G_{w}(\phi) = (2\omega)^{w}T_{w}(\phi). \label{cq}
\end{equation}
It follows  that
\bs
\begin{equation}
<G_{\bar{w}}G_{w^{\prime}}> = (2\omega)^{w}\delta_{w, w^{\prime}}, \;\;\;\;\;\; 
<T_{\bar{w}}T_{w^{\prime}}> = ((2\omega)^{-1})^{w}\delta_{w, w^{\prime}} \label{cm}
\end{equation}
\begin{equation}
G_{wm}G_{nw^{\prime}} = G_{wmnw^{\prime}} + 2 \omega_{m}\delta_{m,n}G_{w}G_{w^
{\prime}} \label{cn}
\end{equation}
\begin{equation}
T_{wm}T_{nw^{\prime}} = T_{wmnw^{\prime}} + X_{mn}T_{w}T_{w^{\prime}} \label{co}
\end{equation}
\es
where (\ref{cm}) and (\ref{cn}) follow from (\ref{am}) and (\ref{cl}) 
respectively, while (\ref{co}) 
follows from (\ref{cn}). The solution of the recursion relation 
(\ref{cn}) is the finite operator product expansion
\begin{equation}
G_{w}G_{w^{\prime}} = \sum_{u} \delta_{w,w_{1}u}\; \delta_{w^{\prime},\bar{u}w_{2}} 
(2\omega)^{u}\;G_{w_{1}w_{2}} \label{ex}
\end{equation}
which is a free-algebraic generalization of a familiar decomposition rule for 
the product of two Chebyshev polynomials (see also the general one-dimensional 
 operator product expansion 
 in Eq. (\ref{cp})). Using (\ref{cq}) in (\ref{ex}), one also obtains the 
 explicit form  (in this case) of the $T_{w}T_{w^{\prime}}$ operator product 
 expansion in (\ref{au}). 

In this case, the interacting Cuntz algebra becomes the Cuntz algebra
\begin{equation}
\{ a_{m}, a^{\dagger}_{m} \} \equiv \{B_{m},B^{\dagger}_{m}\} /{\sqrt{ 
2\omega_{m}}}
\end{equation}
and the infinite dimensional free algebra (\ref{c}) has corresponding 
simplifications due to the simple forms of $G$ and $E$ in (\ref{cs}). We 
mention in particular that  
\begin{equation}
a_{w}a_{w^{\prime}}^{\;\;\;\dagger} = \left\{ \begin{array}{l} 
{\delta_{w,w^{\prime}}}\;\;\; if\;\; [w]=[w^{\prime}] \\ 
a_{u} \;\;\;\;\;\; if\;\; w=uw^{\prime} \\ 
a_{u}^{\;\;\;\dagger} \;\;\;\; if \;\; w^{\prime}=uw \\ 
0 \;\;\;\;\;\;\;\; otherwise   \end{array} \right.
\end{equation}
is the simple form of the infinite dimensional free-algebraic relation
 (\ref{bc}) in this case.

\vskip 1.0cm
\setcounter{equation}{0}
\def\theequation{C.\arabic{equation}}
\boldmath
\noindent{\bf Appendix C. Composite Structure of the Master Constraints}%C
\unboldmath
\vskip 0.5cm
  
Define
\begin{equation}
Q_{mwn} \equiv \pi_{m}G_{w}\pi_{n} +i\pi_{m} 
F_{n\bar{w}}^{\;\;\;\;\;\dagger} -i F_{mw}\pi_{n} + F_{mwn}-F_{mw}F_{n}.
\end{equation}
The master constraints (\ref{aa}) are $Q_{mwn}=0$, but one can show from 
(\ref{cu}) and (\ref{cv}) that
\begin{equation}
Q_{mwnp} = Q_{mwn}B_{p}^{\dagger} + B_{mw}Q_{np} \label{cw}
\end{equation}
without using the constraints. (The cubic terms in $\pi$ on the right simply 
cancel.)

Starting with the two-index $Q$'s
\begin{equation}
Q_{mn} = B_{m}B_{n}^{\dagger} - E_{mn}
\end{equation}
we may iterate (\ref{cw}) to obtain the higher-indexed $Q$'s, for example
\bs
\begin{equation}
Q_{mnp} = Q_{mn}B^{\dagger}_{p} + B_{m}Q_{np}
\end{equation}
\begin{equation}
Q_{mnpq} = (Q_{mn}B_{p}^{\dagger} + B_{m}Q_{np})B_{q}^{\dagger} + 
B_{mn}Q_{pq}
\end{equation}
\es
and one finds more generally that all the $Q$'s are linear in $Q_{mn}$.
 It follows that all the $Q$'s are zero when the first one is set to 
 zero:
\begin{equation}
B_{m}B_{n}^{\dagger} = E_{mn}\;\; \longrightarrow \;\;Q_{mwn} = 0
\end{equation}
and so the set of master constraints (\ref{aa}) contain no new constraints
beyond the first.  

\newpage

\vskip 1.0cm
\setcounter{equation}{0}
\def\theequation{D.\arabic{equation}}
\boldmath
\noindent{\bf Appendix D. Identification of $X(\beta)$}%D
\unboldmath
\vskip 0.5cm

Here we will derive, by simple algebra, the functional relation between 
the generating function 
\begin{equation}
X = X(\beta) = \sum_{w} \beta^{w} X_{w}, \;\;\;\;\;\; 
X_{0} = 0
\end{equation}
and the generating function
\begin{equation}
Z(j) = \sum_{w} j^{w} <\phi^{w}>, \;\;\;\;\;\; Z(0)=1
\end{equation}
of the ordinary planar parts.

Start by rewriting the generator for the polynomials $T_{w}$ as follows
\bs \label{cx}
\begin{eqnarray}
&&\sum_{w} \beta^{w}T_{w} = \frac{1}{1-\beta_{m}\phi_{m} + 
X(\beta)} \\ 
&&= (1+X)^{-1} \frac {1}{1-\phi_{m} \beta_{m}(1+X)^{-1}}  
= (1+X)^{-1} \sum_{w}\; j^{w} \phi^{w}
\end{eqnarray}
\es
where we have made the identification 
\begin{equation}
j_{m} = \beta_{m} (1+X(\beta))^{-1} \label{bn} 
\end{equation}
between the two  sets of free-algebraic sources.
Now multiply (\ref{cx}) on the left by $\phi_{m} \beta_{m}$, take the vev and 
use the definition $X_{mw} = <\phi_{m}T_{w}>$ to get
\begin{equation}
\sum_{m,w} \beta^{mw} \; X_{mw} = \sum_{m,w} j^{mw}  < \phi^{mw} > \;
\end{equation}
which is just
\begin{equation}
X(\beta) = Z(j)-1. \label{bo}
\end{equation}
Combining (\ref{bn}) with (\ref{bo}) we  have
\begin{equation}
Z(j) = 1+X(jZ(j)) \label{dh}
\end{equation}
or alternatively
\begin{equation}
X(\beta) = Z(\beta(1+X(\beta))^{-1}) -1 .
\end{equation}
Following Refs.~\cite{Brez, Cvit}, the relation (\ref{dh}) identifies $X(\beta)$ 
as a generating function of connected planar parts.

Similarly, the relation~\cite{Cvit} 
\begin{equation}
Z(j)=1+X(Z(j)j)
\end{equation}
is obtained by expanding (\ref{cx}) with $(1+X)^{-1}$ on the right and using
 (\ref{ec}).
 
Finally, we can establish the similar relation 
\begin{equation}
\bar{Z}(j) = 1 + \bar{X}(j\bar{Z}(j)), \;\;\;\;\;\; \bar{Z}(j) = 
\sum_{w}j^{\bar{w}}<\phi^{w}> \label{ff}
\end{equation}
for the alternate generating functions $\bar{Z}$ and $\bar{X}$.
To derive this result, start with the relations
\begin{equation}
(1-\beta_{m}\tilde{\phi}_{m} + \bar{X}(\beta))^{-1} = \sum_{w}\beta^{w}
\widetilde{T_{\bar{w}}(\phi)}, \;\;\;\;\;\; 
\widetilde{T_{w}(\phi)} \ra = T_{w}(\phi)\ra .
\end{equation}
These can be derived from Ref.~\cite{us} and (\ref{cj}), (\ref{aq}) and 
(\ref{ar}), and then 
proceed as earlier in this appendix.

\vskip 1.0cm
\setcounter{equation}{0}
\def\theequation{E.\arabic{equation}}
\boldmath
\noindent{\bf Appendix E. Schwinger-Dyson as Null State Ward Identities}%E
\unboldmath
\vskip 0.5cm

There are many free-algebraic forms of the Schwinger-Dyson equations, 
some of which are discussed in Sec.~8.  In this Appendix, we discuss a 
form of the Schwinger-Dyson equations which follows from the Ward 
identities of the infinite-dimensional free algebra.

This development is based on the null states
\begin{equation}
(B^{\dagger\; w} - G_{w}(\phi)) \ra = 0
\end{equation}
which give the null state Ward identities
\begin{equation}
<\phi^{\bar{w}'}(B^{\dagger\; w} - G_{w}(\phi)) > = 0.
\end{equation}
To put these identities in a useful form, we leave the coupling 
constant-dependent $G_{w}(\phi)$ terms as they are and evaluate the 
$B^{\dagger\;w}$ terms as follows:
\bs \label{fd}
\begin{eqnarray}
&&<\phi^{\bar{w}'}\;G_{w}(\phi)> = <\tilde{\phi}^{w'}B^{\dagger\;w}> \\ 
&&= \left\{ \begin{array}{l} \sum_{w \subset w'}\; 
\prod_{\{u_{i}\}=w'/w} <\phi^{u_{i}}> \\
0 \;\;\;when\;\; w \;\; is\; not\;embedded\; in\; w' . \end{array}\right.
\end{eqnarray}
\es
The last form is obtained by writing $B^{\dagger\;w}$ as a product of 
$B^{\dagger}_{m}$'s and moving each to the left using
\begin{equation}
[\tilde{\phi}_{m},B^{\dagger}_{n}] = \delta_{m,n} \ra \la, 
\;\;\;\;\;\; \la B^{\dagger}_{m}=0.
\end{equation}
This procedure shows that the average (\ref{fd}) vanishes unless the 
word $w$ is embedded in the word $w'$, which we write as $w \subset w'$. 
In further detail, $w$ is embedded in $w'$ if the two words can be 
written as
\bs
\begin{eqnarray}
&&w = m_{1}m_{2} \ldots m_{n} \\
w \subset w':&&  w'=u_{1}m_{1}u_{2}m_{2} \ldots u_{n}m_{n}u_{n+1} 
\end{eqnarray}
\es
which defines the ``quotient set'' $\{u_{i}\} = w'/w$ of words 
$u_{i}$ uniquely for each embedding.

As examples of (\ref{fd}) we list
\bs
\begin{eqnarray}
&&<G_{w}> = \delta_{w,0} \label{fe} \\
&&<\phi_{m}G_{n}> = \delta_{m,n} \\
&&<\phi_{m}\phi_{n}\phi_{p}G_{q}> = \delta_{m,q}<\phi_{n}\phi_{p}> + 
\delta_{n,q}<\phi_{m}><\phi_{p}>+\delta_{p,q}<\phi_{m}\phi_{n}> 
\nonumber \\
&&<\phi_{m}G_{np}>=0 \nonumber 
\end{eqnarray}
where (\ref{fe}) was noted in (\ref{ag1}).
\es

\vskip 1.0cm
\setcounter{equation}{0}
\def\theequation{F.\arabic{equation}}
\boldmath
\noindent{\bf Appendix F. Perturbation Theory}%F
\unboldmath
\vskip 0.5cm

We work with the dual basis system (\ref{ey})
and assume that some zeroth-order system has already been solved
\begin{equation}
B^{\dagger}_{m} + E_{mn}^{(0)}(\phi^{(0)})\bar{B}_{n} -G_{m}^{(0)}(\phi^{(0)})=0, 
\;\;\;\;\;\; \phi_{p}^{(0)} = \bar{B}_{p}(1+\bar{X}^{(0)}(B^{\dagger})). \label{dx}
\end{equation}
The general perturbation problem is stated as follows.  Given
\begin{equation}
G_{m}(\phi) = G_{m}^{(0)}(\phi) + \lambda G^{\prime}_{m}(\phi), 
\;\;\;\;\;\;E_{mn}(\phi) = E_{mn}^{(0)}(\phi) + \lambda E^{\prime}_{mn}(\phi)
\end{equation}
we want to solve for the corrections to the connected parts $X_{w}$
\begin{equation}
\bar{X}(B^{\dagger}) - \bar{X}^{(0)}(B^{\dagger}) = \sum_{k=1}^{\infty} \lambda^{k}
\bar{X}^{(k)}(B^{\dagger}) = \sum_{k=1}^{\infty} \lambda^{k} \sum_{w}
X^{(k)}_{\bar{w}}B^{\dagger\; w} 
\end{equation}
order by order in $\lambda$.

We have
\begin{equation}
\phi = \phi^{(0)} + \phi^{\prime} = \phi^{(0)} + \sum_{k=1}\lambda^{k}\bar{B}
\bar{X}^{(k)}(B^{\dagger})
\end{equation}
and  we  subtract (\ref{dx}) from (\ref{bz}) to get the general 
perturbation  equation 
\begin{equation}
[ E_{mn}^{(0)}(\phi)-  E_{mn}^{(0)}(\phi^{(0)})]\bar{B}_{n} - 
[G_{m}^{(0)}(\phi) - G_{m}^{(0)}(\phi^{(0)})] = \lambda [G^{\prime}_{m}(\phi) 
- E^{\prime}_{mn}(\phi)\bar{B}_{n}]. \label{dy}
\end{equation}

If we have oscillators for the zeroth-order problem, this general 
equation simplifies somewhat. But we can work from any zeroth-order 
problem and get the desired results by straightforward algebraic 
computation with (\ref{dy}), remembering that the $B$ operators serve as 
``dummy'' variables, obeying  $\bar{B}_{m}B^{\dagger}_{n} = \delta_{m,n}$.

%****************************

\end{document}